\documentclass[11pt,twoside]{book} 
\usepackage{asp2008s}
\usepackage{times}
\usepackage{lscape}
\usepackage{epsf}

\usepackage{graphicx}
\usepackage{amssymb}
\usepackage{longtable}
\usepackage[figuresright]{rotating}
\usepackage{multirow}

\newcommand{\simless}{\mathbin{\lower 3pt\hbox {$\rlap{\raise 5pt\hbox{$\char'074$}}\mathchar"7218$}}}

\mainmatter

\newlength{\deftabcolsep}
\begin{document}


\title{The Lagoon Nebula and its Vicinity}
\author{N.~F.~H.~Tothill}
\affil{Harvard-Smithsonian Center for Astrophysics,
60 Garden Street, Cambridge, MA 02138, USA\\
and\\
School of Physics, University of Exeter, Stocker Road, Exeter, EX4 4QL, UK}
\author{Marc Gagn\'e}
\affil{Department of Geology and Astronomy, West Chester University,
West Chester, PA 19383, USA}
\author{B.~Stecklum}
\affil{Th\"uringer Landessternwarte Tautenburg, Sternwarte 5,
D-07778 Tautenburg, Germany}
\author{M.~A.~Kenworthy}
\affil{Steward Observatory, University of Arizona, 933 N. Cherry Avenue,
Tucson, AZ 85721, USA}

\begin{abstract}
The Lagoon Nebula is an {\sc Hii} region in the Sagittarius Arm, about
1.3~kpc away, associated with the young (1--3~Myr) open cluster NGC\,6530,
which contains several O stars and several dozen B stars. Lower-mass
cluster members, detected by X-ray and H$\alpha$ emission, and by near-IR
excess, number more than a thousand. Myr-old star formation is traced by
the optically-visible {\sc Hii} region and cluster; observations of
infrared and submillimetre-wave emission, and of optical emission features,
indicate ongoing star formation in several locations across the Lagoon.
The most prominent of these are the Hourglass Nebula and M8\,E.
Submillimetre-wave observations also reveal many clumps of dense molecular
gas, which may form the next generation of stars. The complex structure of
the region has been shaped by the interaction of the underlying molecular
gas with multiple massive stars and episodes of star formation. NGC\,6530
is the oldest component, with the newest stars found embedded in the
molecular gas behind the cluster and at its southern rim. A degree to the
east of the Lagoon, Simeis~188 is a complex of emission and reflection
nebulae, including the bright-rimmed cloud NGC\,6559; the presence of
H$\alpha$ emission stars suggests ongoing star formation.
\end{abstract}

\section{Introduction}

\begin{figure}[thb]
\centering
\includegraphics[draft=False,width=\textwidth]{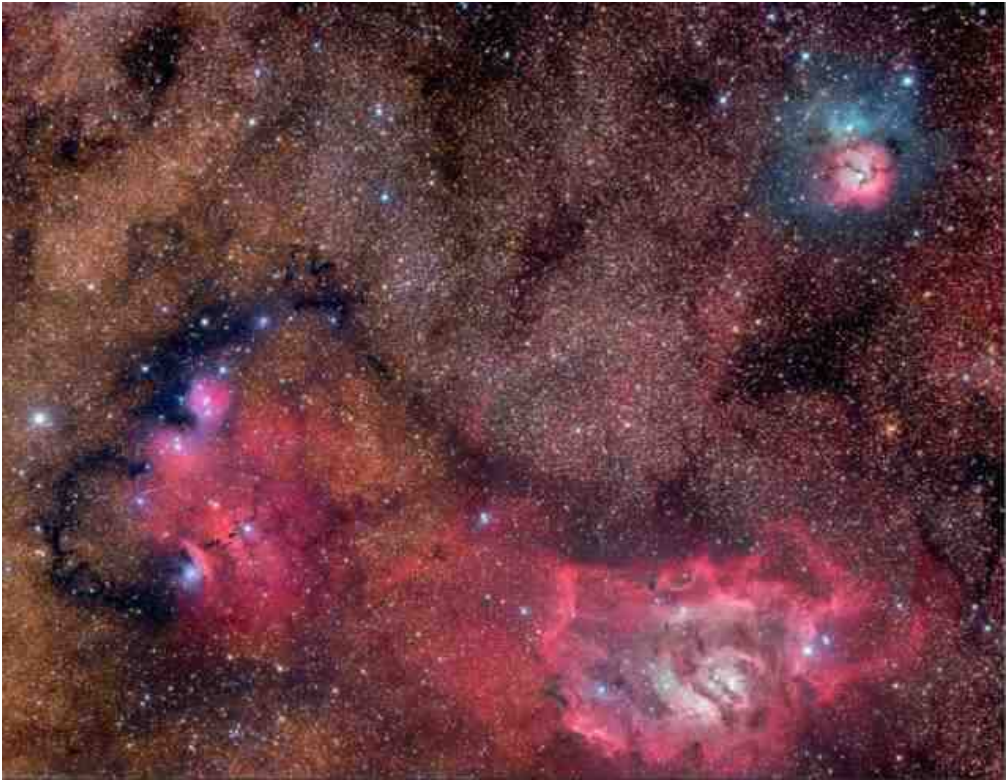}
\caption{A widefield colour image covering M\,8, M\,20 and Simeis 188:
H$\alpha$ emission is red and reflection nebulae are blue; north
is up and east is to the left; field of view (FOV)
$\sim 2.5^\circ\times\sim 2.2^\circ$. Courtesy Gerald Rhemann.}
\label{fig-wide}
\end{figure}

\begin{figure}[thb]
\centering
\includegraphics[draft=False,width=\textwidth]{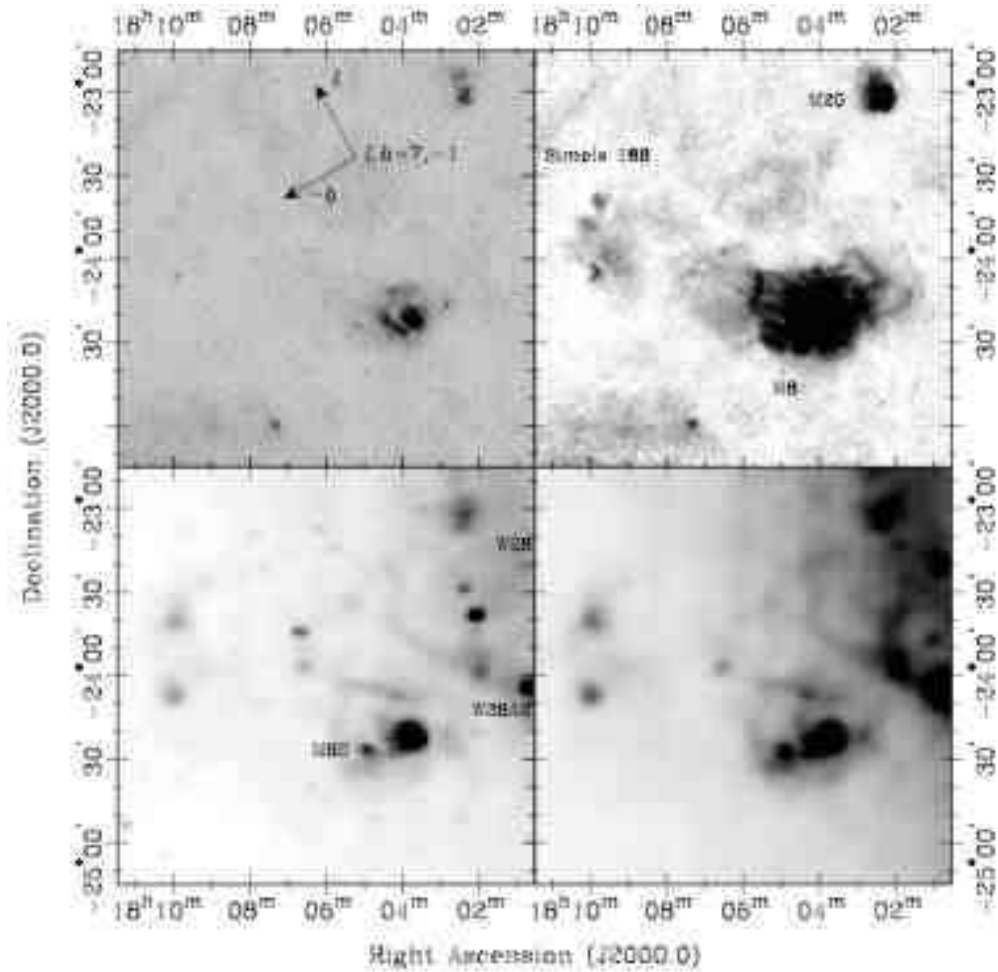}
\caption{A $2.5^\circ\times 2.5^\circ$ field in Sagittarius at 4 wavelengths:
{\it Upper Left:} Digitised Sky Survey (DSS, blue), dominated
by stellar emission; {\it Upper Right:} DSS2 (red), showing H$\alpha$
emission as well; {\it Lower Left:} IRAS 12~$\mu$m;
{\it Lower Right:} IRAS 100~$\mu$m. Arrows denote the directions of
Galactic longitude and latitude (each arrow is 0.5$^\circ$ long).
The Lagoon Nebula and nearby regions are annotated.}
\label{fig-sgr}
\end{figure}

The Lagoon Nebula --- M\,8 --- is the most prominent of a number of
star-forming regions and supernova remnants in the section of the
Sagittarius-Carina Arm lying near our line of sight towards the Galactic
centre (Figs.~\ref{fig-wide} \& \ref{fig-sgr}). Other members include:
The Trifid Nebula (M\,20); the supernova remnant W\,28, near the Trifid at a
kinematic distance of 1.9~kpc \citep{velazquez}; the nearby
optically-invisible {\sc Hii} region W28\,A2 (G\,5.89--0.39), lying between
the Lagoon and Trifid Nebulae at a distance of 2.0~kpc \citep{acord}; and
the complex of nebulae to the east of the Lagoon known as Simeis~188.
The Eagle Nebula (M\,16, NGC\,6611) and M\,17 lie about $10^\circ$ further along
the Sagittarius-Carina arm.

M\,8 consists of a rich open cluster with several O-type stars and a
prominent {\sc Hii} region (about half a degree in diameter), the core of
the cluster superimposed on the eastern half of the {\sc Hii} region.
The {\sc Hii} region is surrounded by bright rims and at least one dark
`elephant trunk' structure; these are most prominent at the southeastern
edge of the {\sc Hii} region. A dark lane splits the optical nebula from
NE to SW (the `Great Rift'); the lack of background stars in the Rift
implies that it is an obscuring dust lane rather than a lack of material,
but it does not show up clearly in submillimetre- or millimetre-wave maps,
either of spectral lines of CO or of the dust continuum; the Rift
presumably has a high enough column density to obscure optical wavelengths
significantly, but not enough to be obvious in emission.

The open cluster is fairly young (a few Myr). It is centred on
18:04:24, --24:21:12 (J2000.0), with a radius of around 30\arcmin,
but a core radius of only $\sim$4\arcmin\ \citep{chen}. It contains
several O stars and about 60 B stars: One of its probable members, the
O4 star \citep[and probable binary,][]{rauw05} 9\,Sagittarii, is the
chief source of ionising flux for the {\sc Hii} region. The western half
of the {\sc Hii} region is concentrated into a bright core which contains
the Hourglass Nebula, a distinctively-shaped window into a compact
{\sc Hii} region with much denser ionised gas than the main body of the
Lagoon Nebula, powered by the young O7 star Herschel~36
\citep[HD~164740,][]{woodward}.

\begin{figure}
\includegraphics[width=\textwidth]{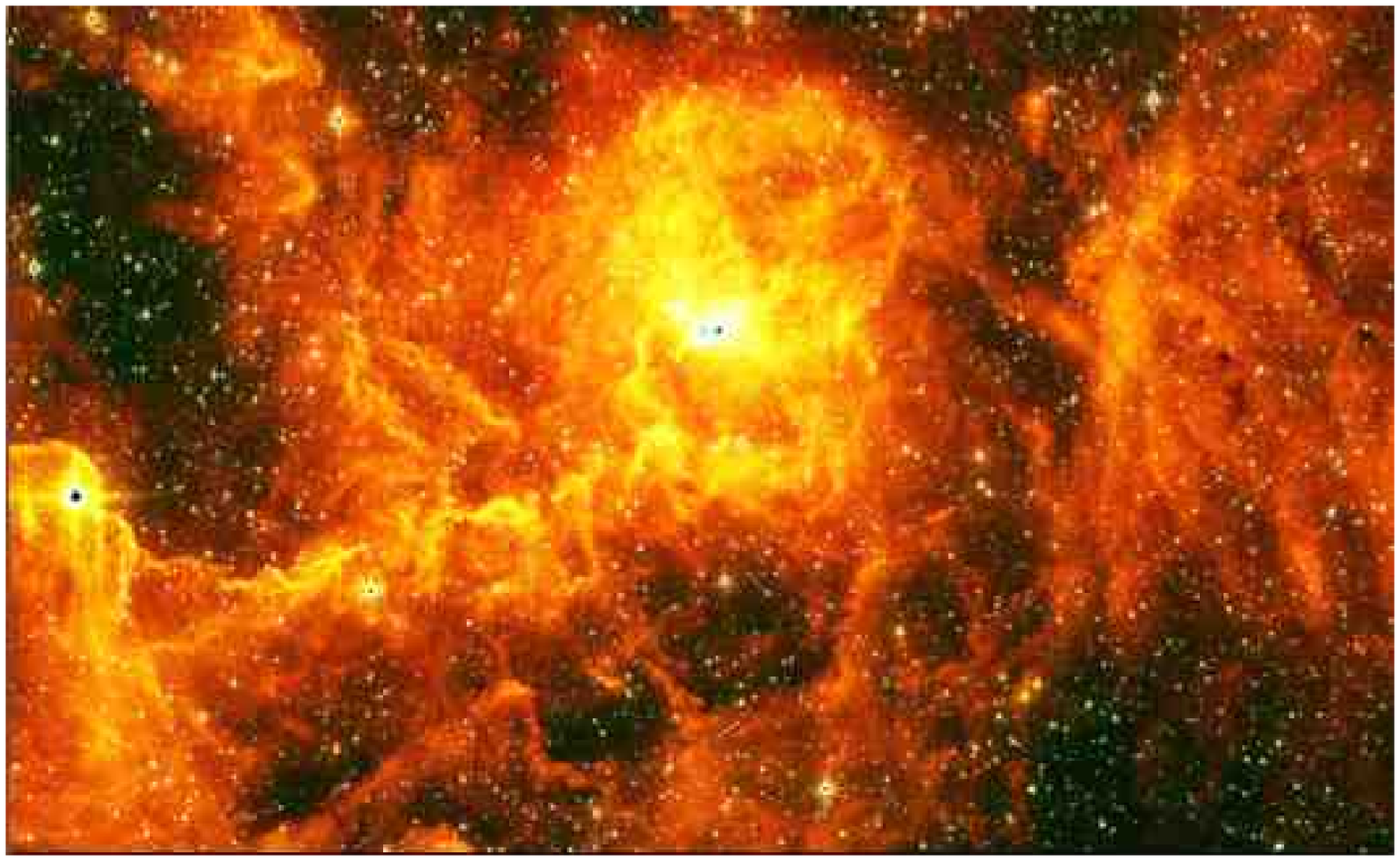}\\[1ex]
\includegraphics[width=\textwidth]{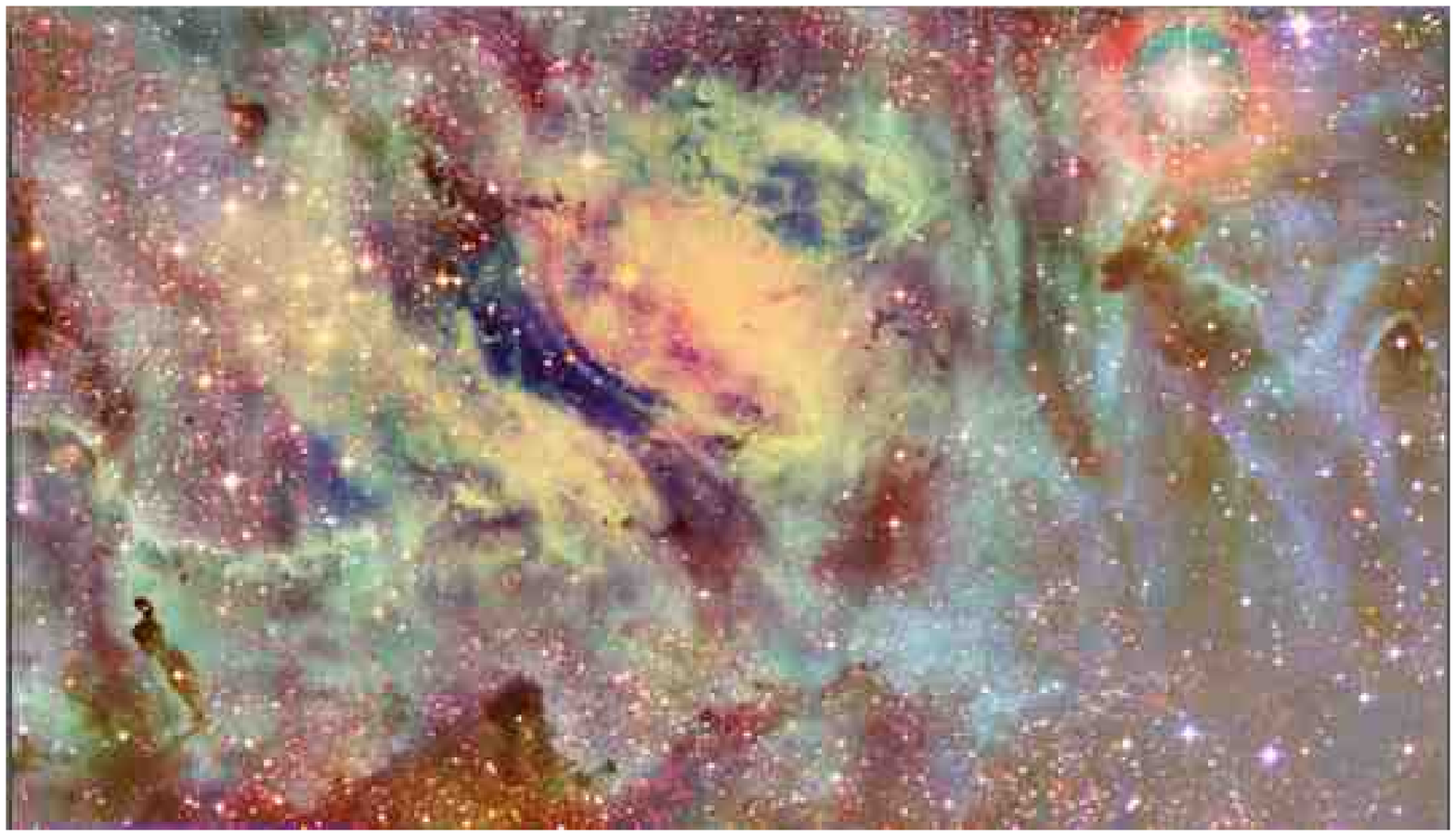}

\caption{{\em Upper:} Four-colour infrared mosaic of the Lagoon Nebula from
{\it Spitzer} IRAC data. Red, orange, green and blue correspond to the 4
IRAC wavelengths (8.0, 5.8, 4.5 \& 3.6~$\mu$m respectively). North is up
and East to the left; FOV is $36^\prime\times 21^\prime$. Based on archive
data from {\it Spitzer} programme 20726, P.I.~J.~Hester.
{\em Lower:} The same field at optical wavelengths, from the Digitised Sky
Survey: Red, green and blue correspond to $I$ and $R$-bands (from DSS2)
and $B$-band (from DSS) respectively. H$\alpha$ emission (in the $R$-band)
appears greenish. M8\,E is only visible in the optical $I$-band, but is
saturated in the infrared. `The Dragon' (Sect.~\ref{sec-over-ism}) is the
prominent `elephant trunk' to the SE of the core of NGC\,6530 (see also
Fig.~\ref{fig-hh}).}
\label{fig-irac}
\end{figure}

Southeast of the cluster core, a structure of bright rims and dark
lanes stretches west to east. CO and dust maps clearly show this to be a
dense molecular cloud; at its eastern end, at least two massive stars are
being formed in the optically-invisible M8\,E cluster which rivals the
brightness of the Hourglass at infrared and submillimetre wavelengths
(see Fig.~\ref{fig-irac}).

The nomenclature of nebulae and clusters in the area is quite unclear,
since the complex has been observed and catalogued since the 17th
century \citep{burnham}. According to the NGC/IC
Project\footnote{http://ngcic.org}, NGC\,6533 refers to the whole nebula
and NGC\,6523 is the bright core of the nebula, lying NW of the Great Rift;
the SE part of the nebula comprises NGC\,6526 in the south and NGC\,6530
in the north; and IC\,1271 and 4678 are small condensations to the east of
the main nebula (IC\,1271 may refer to the O star HD\,165052).

Although Messier referred to it as {\it `amas'}, a cluster
\citep[M\,8,][]{messier}, we will use M\,8 to refer to the whole complex
of stars, {\sc Hii} regions and molecular gas, and `Lagoon Nebula' to be
synonymous. The NE of the region is dominated by the open cluster, so
NGC\,6530 is now always used to refer to the cluster rather than any
surrounding nebulosity. We take NGC\,6533 to refer to the {\sc Hii}
region only, comprising NGC\,6523 and 6526. {\sc Hii} region studies have
generally concentrated on the brighter eastern core (i.e.~NGC\,6523), and
so it is this designation that is found in the literature: For most
practical purposes, NGC\,6533 and 6523 are the same. It is not clear
that IC\,1271 and 4678 refer to real structures, and we will not use these
designations.

Much further east of the Lagoon lies an `R association': a scattering of
smaller reflection and emission nebulae, often known as Simeis~188. Just
to the north of the Lagoon, there is an area of diffuse nebulosity
(Fig.~\ref{fig-sgr}); on purely morphological grounds, this looks like a
diffuse northern extension of NGC\,6533, separated by a dust lane.

This review is structured as follows:
An overview of M\,8 as a whole (Sect.~\ref{sec-over}), broken down into
the main {\sc Hii} region NGC\,6523/6533 (Sect.~\ref{sec-over-hii}), the
young stellar cluster NGC\,6530 (Sect.~\ref{sec-overview-cluster}), its
pre-main-sequence population (PMS, Sect.~\ref{sec-over-ysos}) and the
interstellar medium (Sect.~\ref{sec-over-ism}), followed by an overview
of the distance estimates to M\,8 (Sect.~\ref{sec-distance-nebula}).
Then, major components of the region are discussed in more detail:
NGC\,6530 (Sect.~\ref{sec-ngc6530}), including the PMS stars
(Sect.~\ref{sec-ngc6530-pms}) and reviews of age and distance estimates
(Secs.~\ref{sec-age} \& \ref{sec-distance-cluster}); the Hourglass
Nebula (Sect.~\ref{sec-hg}); M8\,E (Sect.~\ref{sec-m8e}); a few other
candidate star-forming regions (Sect.~\ref{sec-sfrs}); and Simeis\,188
(Sect.~\ref{sec-sim188}). Finally, we briefly discuss the structure and
evolution of the region as a whole (Sect.~\ref{sec-discussion}).

\section{Overview of M\,8}
\label{sec-over}

\subsection{The HII region NGC\,6523/33}
\label{sec-over-hii}

\begin{figure}[thb!]
\centering
\includegraphics[draft=False,width=\textwidth]{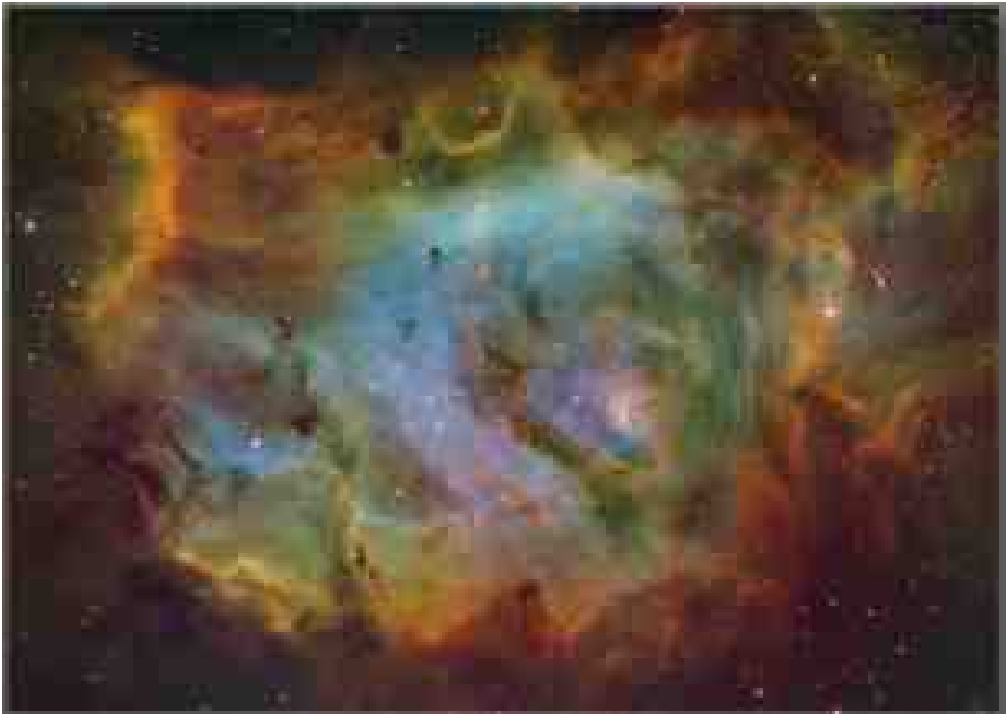}
\caption{Three-colour optical emission-line image of the central part
of M\,8 using interference filters: H$\alpha$ (green), [{\sc Sii}]
(red), and [{\sc Oiii}] (blue). North is up and east is to the left;
FOV is $55^\prime\times 39^\prime$. Courtesy Richard Crisp.}
\label{fig-emission-lines}
\end{figure}

The {\sc Hii} region is about 10~pc in radius and requires about
$10^{51}$ ionising photons per second; 9\,Sgr appears to be its
principal source of ionising radiation, with the binary HD\,165052
\citep[O6.5\,V + O7.5\,V;][]{arias02} contributing as well. Optical
spectroscopy suggests that the bulk of the ionised gas has
electron temperature ($T_e$) about 6000~K \citep{bohuskia}, and electron
density $n_e\sim$500~cm$^{-3}$ increasing to a few $10^3$~cm$^{-3}$
in small condensations and bright rims \citep{bohuskib}. Both density
and temperature increase towards the centre of the {\sc Hii} region.
\citet{lada} identified the star Herschel\,36 as being responsible
for ionising the core of the nebula, an area about 4\arcmin\ across;
it also ionises the Hourglass Nebula, embedded behind this core.
The Hourglass is even hotter and denser, with $n_e$ of 2000--4000~cm$^{-3}$
and $T_e$ of 7000--9000~K, and fluctuations of about 1500~K
\citep{woodward,esteban}. NGC\,6523/33 seems to be a cavity on the front
of a large molecular cloud; the optical emission comes from the working
surface of ionisation fronts moving into a clumpy medium
\citep[Fig.~\ref{fig-emission-lines};][]{elliot, dufour}.
\citet{meaburn71} tentatively detected ionised gas moving at
$\sim$--50~km/s, between us and the nebula, and UV absorptions due to
ionised gas were found at about --30 and --50~km/s \citep{welsh}.
The stellar wind of 9\,Sgr alone would be enough to drive the --50~km/s
shell \citep{welsh}, although other O-stars probably contribute. The
ionised gas in the Lagoon Nebula may be considered as a superposition of
four {\sc Hii} regions: the Hourglass, the core of NGC\,6523 (both powered
by Herschel\,36), the rest of NGC\,6523 and 6533 (ionised by 9\,Sgr), and
the largest and most tenuous component, ionised by HD\,165052
\citep{lyndson,woodward}. More detailed discussion of the {\sc Hii}
region is beyond the scope of this work: The review by \citet{goudis76}
covers radio, optical and IR data.

\subsection{The Open Cluster NGC\,6530}
\label{sec-overview-cluster}

Although the core of NGC\,6530 lies on the line of sight towards a
concentration of molecular gas, the cluster is rather decoupled from the
molecular cloud --- the stars seem to be unobscured, and
\citet{johnson73} showed that the far-IR luminosity of M\,8 is significantly
less than the integrated light of an incomplete census of OB stars in the
cluster; hence, the cluster is not significantly embedded in the molecular
gas. Optical studies all show significant and variable reddening towards
the stars, indicating that there is some interstellar material in front
of the cluster. \citet{mccall} noted that a shell of expanding gas is seen in
absorption against the stars \citep{welsh}, implying that NGC\,6530 lies
within the {\sc Hii} cavity. However, this result is based on
UV spectra of 4 OB stars; membership probabilities for 3 of them are 0.01,
0.08, and 0.39 \citep{vaj}, so this argument is not conclusive.

The surface density of cluster stars, derived from X-ray data
\citep{dam}, shows a compact core about 10\arcmin\ across, surrounded
by a broad extended component to the southeast, south and west. To the
northeast, the surface density of X-ray sources falls off very quickly,
and there is a secondary density peak to the southeast, near the
M8\,E star-forming region. Many optical studies of the cluster, however,
include stars within a field of about a degree, many of which are probable
cluster members, based on proper motion \citep{vaj}.

\begin{figure}
\includegraphics[height=\textwidth,angle=90,draft=False]{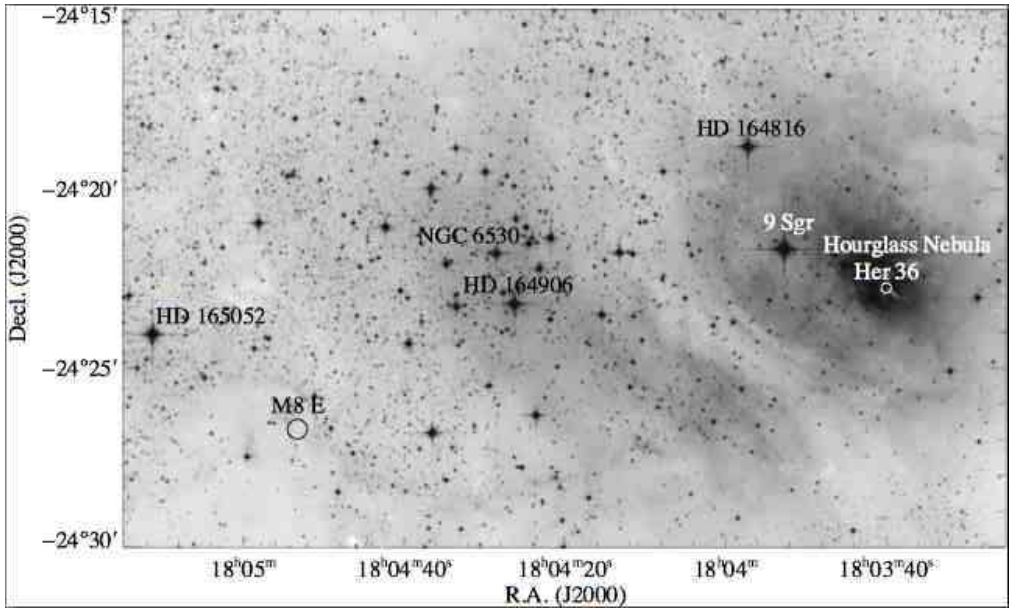}
\caption{25\arcmin$\times$15\arcmin\ portion of the DSS, covering M\,8,
with a square root stretch. The approximate centre of NGC\,6530 is annotated,
as are the Hourglass Nebula, M8\,E, and the brightest stars.}
\label{fig-dss-annot}
\end{figure}

The cluster includes at least 3 O-type stars (see Fig.~\ref{fig-dss-annot}):
The binary HD\,165052, probable binary 9~Sagittarii,
and Herschel~36. HD\,164816 and HD\,164906 (also known as MWC\,280) are
usually classified as O-type \citep[e.g.][]{hiltner}, but have been
reclassified as B3Ve and B0Ve, respectively
\citep{levenhagen}; HD\,164906 may also be binary \citep{roberts}.

\subsection{The Young Stellar Population}

\label{sec-over-ysos}

Studies of the stellar population of NGC\,6530 can be used to estimate
the age and distance of the cluster and thus to characterise the
star formation history of the region, and, by isolating the
pre-main sequence, the young stellar population may be studied.
It is difficult to distinguish the stellar population of NGC\,6530 from
foreground and background stars, since it lies close to the Galactic
plane, and is projected on top of the Galactic bulge. Most studies of
the cluster have been carried out with broadband optical photometry, and
newer studies have extended the photometry into the near-IR.
It is then possible to isolate the cluster in a statistical sense,
e.g.~as a separate population in a colour-magnitude diagram.
Detailed studies of the young stars require individual stars
to be identified as being young. Historically, this has usually involved
looking for H$\alpha$ emission, but modern X-ray imaging with XMM-{\it Newton}
and the ACIS-I camera on {\it Chandra} \citep[][]{rauw,dam} are
a much more efficient way to identify young stars in the cluster:
The X-ray to bolometric flux of T~Tauri stars is at least 100 times
higher than it is for most of the foreground and
background stars which dominate optical and infrared images.
Earlier-type stars
in the cluster are likely to lie on the ZAMS already, and may not show
up in H$\alpha$ or X-ray emission. One of the few possible methods for
identifying individual early-type cluster members is through proper
motion analysis, which relies on the availability of old data.
Fortunately, these early-type stars are rather bright, and thus quite
likely to be visible on old photographic plates \citep[see, e.g.,][]{vaj}.

Quite apart from foreground contamination (in the optical) and
background contamination (in the near-IR), we face the problem that there
are signs of multiple populations of YSOs in the M\,8 region.
Near-IR and X-ray imaging both suggest that there are small dense clusters
associated with the Hourglass Nebula, with M8\,E (both regions of ongoing massive
star formation), and possibly with the Central Ridge.
Disentangling these very young clusters from NGC\,6530 is likely to be
difficult.

\subsection{The Interstellar Medium in the Lagoon Nebula}

\label{sec-over-ism}

\subsubsection{Large-scale structure}

\begin{figure}[tbh]
\plotfiddle{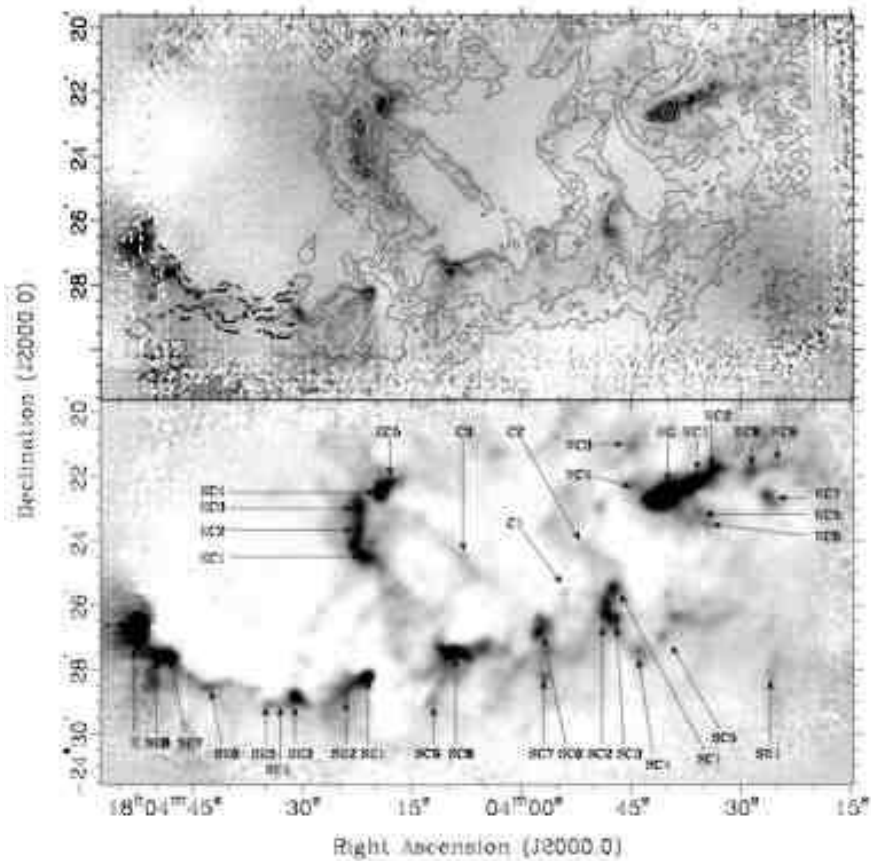}{4.4in}{0}{60.}{60.}{-180}{-10}
\caption{{\it Top:} 450~\micron\ continuum emission (grayscale,
cuts at --1 and 4~Jy\,beam$^{-1}$) with
$^{12}$CO maps overlaid. Dotted contours are at CO 2--1 integrated
intensities of 25 and 50~K\,km\,s$^{-1}$ (black), and 100 and
150~K\,km\,s$^{-1}$ (white);
solid contours are at CO 3--2 integrated intensities of
50 and 100~K\,km\,s$^{-1}$ (black), and 150, 200, 250, 350 and
400~K\,km\,s$^{-1}$(white).
{\it Bottom:} 850~\micron\ continuum emission (grayscale, cuts at
--0.1 and 0.7~Jy\,beam$^{-1}$) with core names annotated.
All data from \citet{thesispaper}}
\label{fig-submm}
\end{figure}

The large-scale structure of the molecular ISM in the Lagoon can be
seen in maps of CO 1--0 \citep{lada} and far-IR continuum emission
\citep{lightfoot}, both with resolutions of 1\arcmin--2\arcmin.
\citeauthor{lada} found 3 bright spots, BS\,1--3: The most prominent
(BS\,1) corresponds to the inner core of the {\sc Hii} region, centred on
the Hourglass (clump HG and surroundings in Fig.~\ref{fig-submm});
BS\,2 (SC\,8 and surroundings in Fig.~\ref{fig-submm}) lies at the southern
edge of the nebula, coincident with the `South Eastern Bright Rim'
(an ionisation front eroding a concentration of gas). BS\,3 lies on the
western edge of the core of NGC\,6530: The continuum map shows it as
an elongated structure, or `eastern bar' (which we call the `central ridge',
clumps EC\,1--5 in Fig.~\ref{fig-submm}). \citeauthor{lada} also found a
CO cloud with significantly redder velocity (28 km/s) lying between the
core of NGC\,6530 and the bright rim linking the Lagoon Nebula to M8\,E.
A near-IR extinction map with comparable resolution shows very similar
structure \citep{dam06}.

\subsubsection{Small-scale structure}
Maps of the molecular ISM (submillimetre continuum and low-$J$ CO)
covering a similar area to those of \citet{lada}, but with resolution
of order 10\arcsec\ were published by \citet{thesispaper}; some of the
maps are shown in Figure~\ref{fig-submm}. The molecular emission is
clearly dominated by the Hourglass and M8\,E, but many other concentrations
become visible. BS\,2 (and the SEBR) can now be resolved as a clump of
gas (SC\,8) with a very sharp transition from the molecular gas to the
ionised gas of the {\sc Hii} region, consistent with the picture of a
structure dominated by the progess of an ionisation front. The Central
Ridge (BS\,3), by contrast, shows up as a complex extended structure of
molecular material (EC\,1--5) running approximately N--S, without the
sharp edges seen in the former. It is possible that this complex has a
similar ionisation structure, but that we observe it face on, the
ionisation front covering the whole of the complex, rather than seeing a
cross-section as with the southern rim. In the Central Ridge, the brightest
continuum emission is seen in a compact condensation at the
northern end, while the CO emission peaks further south along the extended
ridge. This may reflect structural differences, with a dense but relatively
cool clump at the northern end of the warmer, more diffuse ridge.
The lower-resolution continuum data showed a western extension of the
Hourglass emission \citep{lightfoot}, and the newer data resolve this
into a series of clumps running WNW from the Hourglass, also seen in
molecular data \citep{white97}.

The Hourglass Nebula and M8\,E are known to be regions of very recent or
ongoing star formation. The Central Ridge lies close to a cluster
of X-ray sources, suggesting that it too is a locus of star formation
within the Lagoon Nebula. The ionisation-dominated structure of BS2/M8\,SC8
should also be considered a candidate for star formation,
possibly triggered by the compression of the molecular gas by the ionisation
front.

In addition, there are many other molecular clumps scattered throughout the
nebula. Most of them are not particularly dense, but \citet{thesispaper}
suggested that star formation might be triggered in them by the effect of
ionisation fronts (based on the assumed ionising flux from 9\,Sgr and
Herschel~36). \citet{brand} studied a dark `elephant trunk' structure (which
they call `The Dragon') lying south of the ridge that connects M8\,SE1 to
M8\,E (see Fig.~\ref{fig-irac}), using stellar intensity profiles to model
the cloud as a $1\times 0.1$~pc structure of density a few $10^3$~cm$^{-3}$,
and total mass $>9$~M$_\odot$. The Dragon can also be seen in the 850~$\mu$m
continuum \citep{thesispaper} as a very faint structure just visible over
the noise. Based purely on its position, it would seem to lie outside and in
front of the arc of dense gas seen in the submillimeter continuum. The
question of whether or not the Dragon is associated with star formation
remains open.

In the course of a larger survey of high-mass protostar candidates,
\citet{beltran} mapped the M\,8 area at 1.3~mm wavelength with
a beam about twice the size of that of \citet{thesispaper}.
The structure seen in their map is similar, and their published brightnesses
are comparable to those seen at 1.3~mm in the earlier study.
Table~\ref{tab-submm}
lists the clumps seen in the two studies, cross-referenced where possible.
\citeauthor{beltran} also examined the MSX images, and found
that all of the clumps in their map are associated with mid-IR emission,
almost all of them with point sources.

The clump masses derived by \citeauthor{beltran} are systematically
higher than those from \citet{thesispaper}, by up to 2 orders of magnitude
(averaging about a factor of 10). This very large discrepancy can be
explained by a combination of factors: \citet{beltran} often find
larger fluxes from a clump--- the most extreme example is the Hourglass
clump (M8\,HG/Clump~1) with 1.3~mm fluxes of 2.6 and 14~Jy measured by
\citeauthor{thesispaper} and \citeauthor{beltran} respectively.
These larger fluxes reflect the larger beamsize of the later
observations and the use of {\sc CLUMPFIND} to decompose the maps into
clumps. {\sc CLUMPFIND} packs all the flux in the map above a certain
limit into the various clumps it finds, whereas decomposition into
gaussian clumps (used by \citeauthor{thesispaper}) tends to
leave residual flux. There are further discrepancies, due to differences
in the conversion from continuum flux to gas mass between the two papers.
For M\,8, \citeauthor{beltran} use $M/S_{1.3} = 99$, whereas
\citeauthor{thesispaper} use a range of about $10-30$. This difference
arises from assumptions about the mm-wave opacity of dust
(1~cm$^2$~g$^{-1}$ against 1.3~cm$^2$~g$^{-1}$), the temperature of
the dust, and the distance to M\,8. \citeauthor{beltran} assume
a temperature of 30~K for all clumps, while \citeauthor{thesispaper}
derive individual clump temperatures from CO observations. Since the
mass/flux ratio is inversely proportional to the temperature (in the
Rayleigh-Jeans limit), this is not a large effect, but can reach almost
a factor of two: for the Hourglass, \citeauthor{thesispaper} adopted a
temperature of 48~K. By comparing the velocity of CS emission to a
Galactic rotation curve, \citeauthor{beltran} estimated the distance to
M\,8 to be 3.1~kpc, compared to the 1.7~kpc assumed by
\citeauthor{thesispaper} Since the mass-to-flux ratio is proportional
to the square of the distance adopted, this produces a discrepancy in
mass of more than a factor of 3. The combination of all these discrepancies
is enough to account for the differences in the mass estimates.
The distance adopted by \citeauthor{beltran} is inconsistent with the
accepted range (see later in this review), and the distance adopted
by \citeauthor{thesispaper} is also a bit further than the most likely
current estimate of 1.3\,kpc. The gas masses quoted in
Table~\ref{tab-submm} have been rescaled to a fiducial distance of
1.3\,kpc; remaining discrepancies reflect the uncertainties in the
absolute derivation of mass from dust emission.

\begin{table}[tbp]
\caption{Submillimetre Molecular Gas Features in M\,8}
\label{tab-submm}
\smallskip
\begin{center}
{\small
\begin{tabular}{lcccccccc}
\tableline
\noalign{\smallskip}
Name\tablenotemark{a} & No.\tablenotemark{b} &
R.A.\tablenotemark{c} & Dec.\tablenotemark{c}
& $M$\tablenotemark{a,d} & $M$\tablenotemark{b,d}
& $T$\tablenotemark{e} & $D$\tablenotemark{b} & $D$\tablenotemark{a}\\
 & & (J2000.0) & (J2000.0) & ($M_\odot$) & ($M_\odot$) & (K) & (pc) & (pc) \\
\noalign{\smallskip}
\tableline
\noalign{\smallskip}
M8HG  & 1  & $18$:$03$:$40.7$ & $-24$:$22$:$40$ & $7.8$  & $241$  & $48$ &
$1.0$ & $0.2$ \\
M8WC1 &    & $18$:$03$:$36.6$ & $-24$:$22$:$14$ & $5.5$  &        & $30$ &
      & $0.2$ \\
M8WC2 &    & $18$:$03$:$33.7$ & $-24$:$21$:$49$ & $9.7$  &        & $24$ &
      & $0.3$ \\
M8WC3 &    & $18$:$03$:$44.8$ & $-24$:$21$:$23$ & $1.3$  &        & $25$ &
      & $0.2$ \\
M8WC4 &    & $18$:$03$:$44.6$ & $-24$:$22$:$16$ & $0.6$  &        & $36$ &
      & $0.3$ \\
M8WC5 &    & $18$:$03$:$35.9$ & $-24$:$23$:$10$ & $1.3$  &        & $31$ &
      & $0.2$ \\
M8WC6 &    & $18$:$03$:$34.6$ & $-24$:$23$:$25$ & $4.5$  &        & $16$ &
      & $0.2$ \\
M8WC7 & 22 & $18$:$03$:$26.2$ & $-24$:$22$:$34$ & $7.3$  & $2.1$  & $20$ &
$0.2$ & $0.2$ \\
M8WC8 &    & $18$:$03$:$28.5$ & $-24$:$21$:$50$ & $10.1$ &        & $16$ &
      & $0.2$ \\
M8WC9 &    & $18$:$03$:$25.3$ & $-24$:$21$:$39$ & $4.9$  &        & $13$ &
      & $0.2$ \\
M8SW1 & 21 & $18$:$03$:$25.8$ & $-24$:$28$:$11$ & $4.3$  & $3.5$  & $15$ &
$0.3$ & $0.2$ \\
M8EC1 & 12 & $18$:$04$:$21.6$ & $-24$:$24$:$27$ & $4.3$  & $23.6$ & $25$ &
$0.5$ & $0.2$ \\
M8EC2 &    & $18$:$04$:$22.5$ & $-24$:$23$:$25$ & $3.0$  &        & $35$ &
      & $0.2$ \\
M8EC3 & 13 & $18$:$04$:$22.4$ & $-24$:$22$:$57$ & $4.2$  & $14.6$ & $36$ &
$0.4$ & $0.2$ \\
M8EC4 & 6  & $18$:$04$:$19.2$ & $-24$:$22$:$26$ & $8.0$  & $15.5$ & $30$ &
$0.2$ & $0.2$ \\
M8EC5 & 5  & $18$:$04$:$18.0$ & $-24$:$22$:$05$ & $3.6$  & $18.8$ & $31$ &
$0.2$ & $0.2$ \\
M8E   & 2  & $18$:$04$:$52.6$ & $-24$:$26$:$35$ & $20.0$ & $127$  & $29$ &
$0.7$ & $0.2$ \\
M8SE1 & 10 & $18$:$04$:$21.6$ & $-24$:$28$:$17$ & $6.4$  & $49.5$ & $23$ &
$0.8$ & $0.2$ \\
M8SE2 &    & $18$:$04$:$24.4$ & $-24$:$28$:$39$ & $1.2$  &        & $31$ &
      & $0.2$ \\
M8SE3 & 11 & $18$:$04$:$31.1$ & $-24$:$28$:$53$ & $7.3$  & $25.0$ & $21$ &
$0.7$ & $0.3$ \\
M8SE4 &    & $18$:$04$:$32.9$ & $-24$:$29$:$08$ & $0.5$  &        & $21$ &
      & $0.1$ \\
M8SE5 & 18 & $18$:$04$:$34.4$ & $-24$:$29$:$05$ & $2.8$  & $6.2$  & $23$ &
$0.2$ & $0.2$ \\
M8SE6 & 19 & $18$:$04$:$43.1$ & $-24$:$28$:$32$ & $3.8$  & $7.2$  & $22$ &
$0.4$ & $0.2$ \\
M8SE7 & 7  & $18$:$04$:$48.5$ & $-24$:$27$:$33$ & $7.2$  & $66.2$ & $30$ &
$0.9$ & $0.2$ \\
M8SE8 &    & $18$:$04$:$50.5$ & $-24$:$26$:$59$ & $3.2$  &        & $29$ &
      & $0.2$ \\
M8SC1 &    & $18$:$03$:$47.5$ & $-24$:$25$:$31$ & $1.8$  &        & $38$ &
      & $0.2$ \\
M8SC2 & 3  & $18$:$03$:$48.1$ & $-24$:$26$:$18$ & $7.2$  & $41.7$ & $28$ &
$0.3$ & $0.3$ \\
M8SC3 &    & $18$:$03$:$47.3$ & $-24$:$26$:$33$ & $4.7$  &        & $16$ &
      & $0.2$ \\
M8SC4 &    & $18$:$03$:$43.9$ & $-24$:$27$:$28$ & $4.1$  &        & $21$ &
      & $0.3$ \\
M8SC5 &    & $18$:$03$:$40.7$ & $-24$:$27$:$59$ & $1.3$  &        & $29$ &
      & $0.2$ \\
M8SC6 & 8  & $18$:$03$:$57.6$ & $-24$:$26$:$22$ & $1.9$  & $11.1$ & $21$ &
      & $0.2$ \\
M8SC7 &    & $18$:$03$:$57.0$ & $-24$:$28$:$12$ & $1.0$  &        & $25$ &
      & $0.2$ \\
M8SC8 & 9  & $18$:$04$:$09.5$ & $-24$:$27$:$30$ & $7.5$  & $57.9$ & $37$ &
$0.9$ & $0.3$ \\
M8SC9 &    & $18$:$04$:$12.2$ & $-24$:$28$:$58$ & $2.9$  &        & $28$ &
      & $0.3$ \\
M8C1  &    & $18$:$03$:$54.1$ & $-24$:$25$:$40$ & $1.5$  &        & $16$ &
      & $0.2$ \\
M8C2  &    & $18$:$03$:$51.2$ & $-24$:$24$:$14$ & $0.2$  &        & $41$ &
      & $0.2$ \\
M8C3  &    & $18$:$04$:$08.2$ & $-24$:$24$:$30$ & $0.9$  &        & $21$ &
      & $0.1$ \\
      & 4  & $18$:$03$:$49.8$ & $-24$:$26$:$07$ &        & $29.4$ &      &
$0.2$ &       \\
      & 14 & $18$:$04$:$19.1$ & $-24$:$23$:$27$ &        & $6.9$  &      &
$0.1$ &       \\
      & 15 & $18$:$04$:$54.8$ & $-24$:$25$:$27$ &        & $15.5$ &      &
$0.4$ &       \\
      & 16 & $18$:$03$:$36.9$ & $-24$:$26$:$07$ &        & $5.8$  &      &
$0.3$ &       \\
      & 17 & $18$:$04$:$17.9$ & $-24$:$24$:$15$ &        & $8.5$  &      &
$0.4$ &       \\
      & 19 & $18$:$04$:$23.7$ & $-24$:$22$:$47$ &        & $7.2$  &      &
$0.4$ &       \\
\noalign{\smallskip}
\tableline
\noalign{\smallskip}
\multicolumn{9}{l}{$^a$ \citet{thesispaper};~~$^b$ \citet{beltran}} \\
\multicolumn{9}{l}{$^c$ taken from \citet{thesispaper} unless it occurs
only in \citet{beltran}} \\
\multicolumn{9}{l}{$^d$ gas masses revised to assumed distance of 1.3~kpc} \\
\multicolumn{9}{l}{$^e$ derived from CO data \citep{thesispaper}} \\
\end{tabular}
}
\end{center}
\end{table}


\subsubsection{Optical Features in the Interstellar Medium}

\begin{figure}[tbh]

\plotfiddle{HH896_897_arXiv.eps}{4.7in}{0}{35.}{35.}{-185}{0}
\plotfiddle{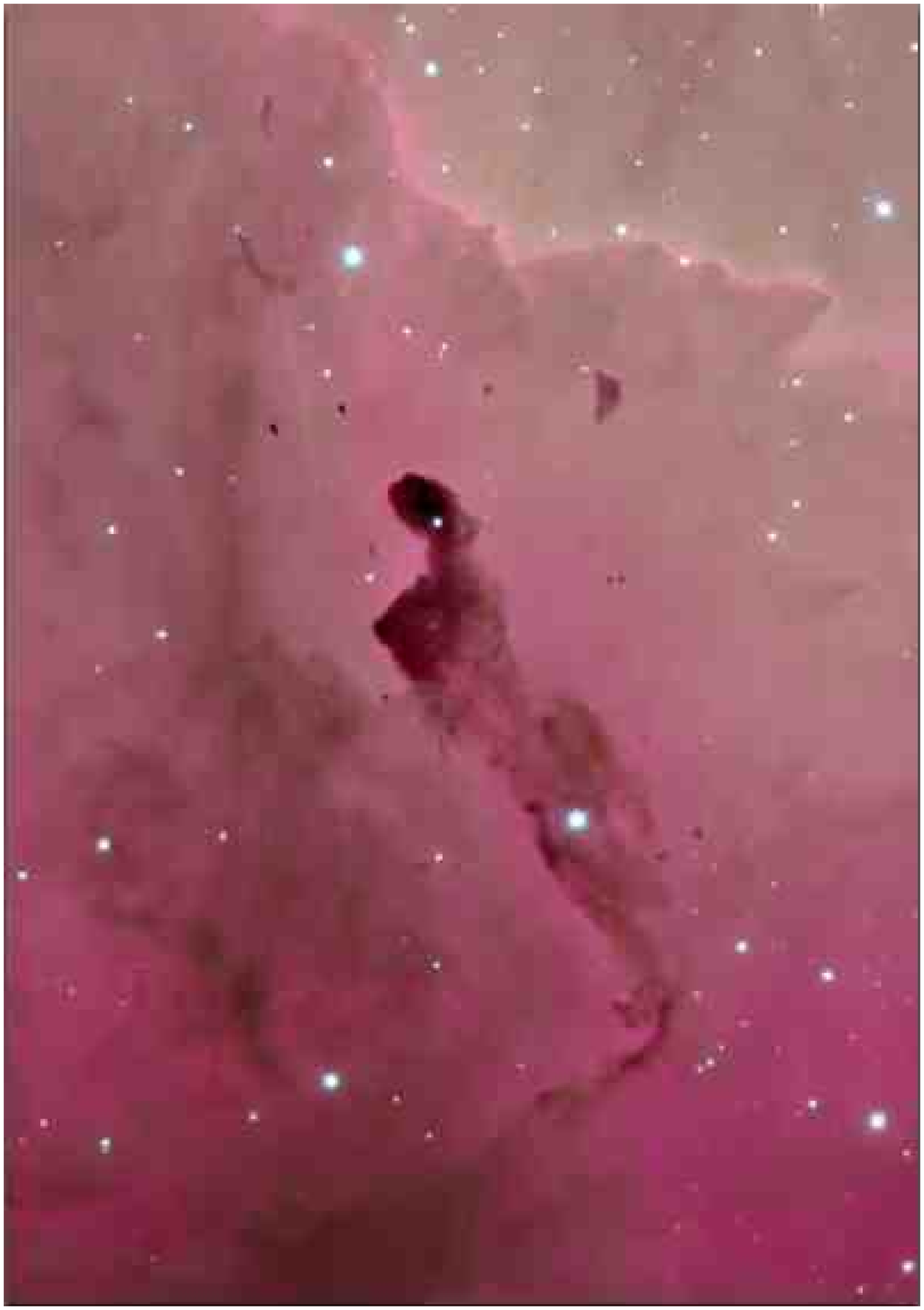}{0pt}{0}{30.}{30.}{0}{60}


\caption{Optical ISM features in the Lagoon: {\em Left:}
[{\sc Sii}] image of HH\,896/897: HH\,896 is at the top, and HH\,897
at the bottom. Axes are in arcseconds, and numbers refer to T~Tauri stars
identified by \citet{arias07}. IRAS 18014--2428 is embedded in the
bright-rimmed clump \citep[M8\,SE3,][]{thesispaper}. From \citet{barba}.
{\em Right:} Broadband colour optical image of `The Dragon', an elephant
trunk lying in front of the SE rim. North is up and East is to the left;
FOV is $\sim 5^\prime\times\sim 7^\prime$. Excerpt from a mosaic obtained at
the CFHT by Jean-Charles Cuillandre.}
\label{fig-hh}
\end{figure}

Optical images of M\,8 reveal a region rich in optical ISM features, such
as Bok globules, bright-rimmed clouds, proplyd-like objects,
Herbig-Haro (HH) objects, etc.~(Table~\ref{tab-optical}), many of which
are generally associated with star formation.

The presence of dark markings against the nebular background was noted
by \citet{barnard08} and \citet{duncan}\footnote{Duncan claims that the
name `Lagoon Nebula' refers to these structures.}. Out of 23 dark areas,
\citet{bok} consider 16 to be ``true globules'', being regular and
round, with diameters ranging from 6\arcsec\ to 1\arcmin\ --- mainly in
the 10\arcsec--30\arcsec\ range.

\citet{sugitani} examined the ESO Schmidt atlas, finding two bright-rimmed
clouds in M\,8 (and one in Simeis~188) associated with IRAS sources, which
they consider to be good candidates for sites of star formation caused by
radiatively-driven implosion. The first (near IRAS\,17597--2422) has a
tightly-curved bright rim, about 200\arcsec\ across; the second
(near IRAS\,18012--2407) has a gentler curvature, the rim being about
280\arcsec\ by 120\arcsec. These clouds are significantly larger than
the globules found by \citeauthor{bok}, which may simply reflect different
decisions as to what constitutes a cloud or globule within the structure
of the interstellar medium. A lack of accurate IRAS fluxes (probably due to
confusion) prevented \citeauthor{sugitani} from classifying the associated
IRAS point sources. These bright-rimmed clouds lie well to the north and west
of the main emission region of M\,8, where they have not been studied by
other authors, although \citet{lada} note a `bright rim' to the north of
NGC\,6530, which may well be the same as the northern bright-rimmed cloud of
\citeauthor{sugitani}. These structures raise the possibility of ongoing
star formation significantly outside the core of the complex.

At least one proplyd has been found in the Lagoon Nebula, around
the B star at the centre of the {\sc UCHii} region G\,5.97--1.17
\citep{stecklum}. \citet{demarco} found 4 proplyd-like objects in an
HST image of a small part of M\,8, but none of them have visible central
stars, suggesting that they are not true proplyds. Near the Hourglass,
there are two T~Tauri stars with bow-shocks around them, pointing towards
Herschel~36, but these are more likely produced by collisions between the
stellar winds \citep{arias05}.

The first HH object noted in the Lagoon Nebula \citep[HH\,213,][]{bo}
lies $>10$\arcmin\ W of the Hourglass. Recent wide-field narrow-band
optical imaging has revealed HH objects around the Hourglass \citep{arias05}
and the southern and southeastern bright rims \citep{barba}. One of these,
HH\,894, has an axis pointing towards the PMS Fe/Ge star ABM\,22, which lies
at the bright rim of the molecular clump SC\,8; HH\,896 and
897, meanwhile,
 seem to come straight out of another molecular clump, SE\,3
\citep[Fig.~\ref{fig-hh};][]{arias07}. Three more PMS stars --- ABM\,21,
27 and 29 --- are found within knots of high-excitation gas \citep{arias07}.
Only a small fraction of the Lagoon Nebula has been studied in this way,
so there may be many more HH objects to be found.

\begin{table}[tbp]
\caption{Optical Interstellar Medium Features in M\,8}
\label{tab-optical}
\smallskip
\begin{center}
{\small
\begin{tabular}{l@{\hskip8pt}c@{\hskip6pt}cc@{\hskip-6pt}}
\tableline
\noalign{\smallskip}
Name  & R.A.(J2000.0) & Dec.(J2000.0) & Notes \\
\noalign{\smallskip}
\tableline
\noalign{\smallskip}
HH 213           & 18:02:30.5 & --24:17:12 & HH Object\tablenotemark{a}\\
IRAS 17597--2422 & 18:02:51.6 & --24:22:08
                                    &Bright-rimmed cloud\tablenotemark{b}\\
HH 869           & 18:03:35.7 & --24:22:30 &   HH Object\tablenotemark{c}\\
HH 868           & 18:03:36.0 & --24:22:49 &   HH Object\tablenotemark{c}\\
HH 867           & 18:03:36.8 & --24:22:33 &   HH Object\tablenotemark{c}\\
ABMMR-CG         & 18:03:36.9 & --24:23:58 &     Globule\tablenotemark{c}\\
G 5.97--1.17     & 18:03:40.5 & --24:22:44 &
                                 {\sc UCHii} \& Proplyd\tablenotemark{d}\\
ABMMR 334-BS     & 18:03:40.5 & --24:23:32
                                  & Bow-shock around star\tablenotemark{c}\\
ABMMR 349-BS     & 18:03:40.7 & --24:23:16 &
                                    Bow-shock around star\tablenotemark{c}\\
HH 870           & 18:03:41.4 & --24:23:25 &    HH Object\tablenotemark{c}\\
NGC\,6530 PLF\,4  & 18:03:44.2 & --24:19:23 &
                                     Proplyd-like feature\tablenotemark{e}\\
NGC\,6530 PLF\,3  & 18:03:44.8 & --24:19:47 &
                                     Proplyd-like feature\tablenotemark{e}\\
NGC\,6530 PLF\,2  & 18:03:45.3 & --24:19:45 &
                                     Proplyd-like feature\tablenotemark{e}\\
NGC\,6530 PLF\,1  & 18:03:45.6 & --24:19:41 &
                                     Proplyd-like feature\tablenotemark{e}\\
HH 895\,A        & 18:03:57.2 & --24:28:04 &    Bow shock\tablenotemark{f}\\
HH 895\,B        & 18:03:59.2 & --24:27:53 &
                                           Knotty filament\tablenotemark{f}\\
B 296            & 18:04:04.4 & --24:32:00 &
                              Barnard dark nebula, 6\arcmin$\times$1\arcmin\\
HH 893\,B        & 18:04:06.0 & --24:24:47 &{\sc Sii} knot\tablenotemark{f}\\
HH 893\,A        & 18:04:06.1 & --24:24:46 &{\sc Sii} knot\tablenotemark{f}\\
ABM 21/SCB 418   & 18:04:10.3 & --24:23:23 &
                                          PMS star in knot\tablenotemark{g}\\
ABM 27/SCB 486   & 18:04:16.0 & --24:18:46 &
                                          PMS star in knot\tablenotemark{g}\\
ABM 29/SCB 495   & 18:04:16.4 & --24:24:39 &
                                          PMS star in knot\tablenotemark{g}\\
IRAS 18012--2407 & 18:04:16.8 & --24:06:59 &
                                       Bright-rimmed cloud\tablenotemark{b}\\
HH 894\,C        & 18:04:17.7 & --24:26:16 &      filament\tablenotemark{f}\\
HH 894\,B        & 18:04:22.0 & --24:25:55 &         knots\tablenotemark{f}\\
HH 894\,A        & 18:04:22.9 & --24:25:52 &     bow shock\tablenotemark{f}\\
HH 896\,A        & 18:04:28.6 & --24:26:38 &     bow shock\tablenotemark{f}\\
HH 896\,B        & 18:04:29.8 & --24:26:57 &     bow shock\tablenotemark{f}\\
HH 896\,C        & 18:04:30.4 & --24:26:20 &faint bow shock\tablenotemark{f}\\
HH 897\,C        & 18:04:30.9 & --24:28:59 &arcs and knots\tablenotemark{f}\\
HH 897\,B        & 18:04:31.0 & --24:29:33 &     filaments\tablenotemark{f}\\
HH 897\,A        & 18:04:31.4 & --24:30:27 &knotty bow shock\tablenotemark{f}\\
B 88             & 18:04:35.0 & --24:06:52 &
                             Barnard dark nebula, 2\arcmin$\times$30\arcsec\\
`The Dragon'     & 18:04:45.2 & --24:30:00
                                     & Dark elephant trunk\tablenotemark{h}\\
B 89             & 18:04:59.8 & --24:21:50 &
                            Barnard dark nebula, 30\arcsec$\times$30\arcsec\\
\noalign{\smallskip}
\tableline
\noalign{\smallskip}
\multicolumn{4}{l}{\parbox{0.9\textwidth}{\footnotesize
    $^a$ \citet{bo};~~$^b$ \citet{sugitani};
    $^c$ \citet{arias05};~~$^d$ \citet{stecklum}
    $^e$ \citet{demarco}; positions taken directly from FITS WCS
    $^f$ \citet{barba};~~$^g$ \citet{arias07};~~$^h$ \citet{brand} }}\\
\end{tabular}
}
\end{center}
\end{table}

\subsection{The Distance to the Lagoon Nebula}

\label{sec-distance-nebula}
Determination of the distance to M\,8 is based on the distance to NGC\,6530.
The physical association of NGC\,6530 with M\,8 is based not just on the
fact that the two lie along the same line of sight; the reddening of the
cluster stars is small but significant, suggesting that they
are neither background nor foreground objects \citep{vda}. The
radial velocities of the cluster stars are also fairly close to that
of the nebula. The various distance determinations for NGC\,6530 are
reviewed in detail in Sect.~\ref{sec-distance-cluster}: To summarise the
discussion, we recommend a distance of 1.3~kpc, with an error of maybe
0.1~kpc. However, there are also several distance estimates of 1.8~kpc.
The discrepancy of about 30\% between the two is very significant: Many
derived quantities depend on $d^2$, and may therefore suffer systematic
errors of $>50\%$.

Recently, distances have been derived to star-forming regions by measuring
the parallax of 22\,GHz H$_2$O masers with VLBI \citep[e.g.][]{vera}.
Despite multiple searches, no 22\,GHz masers have been found towards M\,8,
but M8\,E contains very strong methanol masers. It may therefore be
possible to measure the distance to the sites of star formation in M\,8
directly.

\citet{humphreys} derived a mean distance to Sgr~OB1 of
$1.8\pm 0.1$~kpc, but adopted a distance of 1.6~kpc. \citet{gg70a} cited
a distance of 1.6~kpc, although their fitted trace of the
Sagittarius-Carina spiral arm \citep{gg70b} passes 1.9~kpc away from the
Sun at a
Galactic longitude of 6\deg. The spiral arm trace is based, in this
longitude range, on a group of bright optical {\sc Hii} regions (including
M\,8) with a mean distance of 2.2~kpc.
A distance estimate of 1.3~kpc suggests that NGC\,6530, and hence M\,8, are
located some distance in front of the Sagittarius-Carina arm.

\section{NGC\,6530}

\label{sec-ngc6530}

\subsection{The Main-sequence Population}

\citet{walker} published \ubv\ photometry for 118 stars, concentrated
towards the core of the cluster, as well as a smaller list of variable
stars. \citet{kilambi} extended Walker's list to a total of 319
stars\footnote{the combined Walker-Kilambi numbering system is the one
adopted by WEBDA, and is referred to as `Walker' in this work},
largely to the west and north, and \citet{vaj} used proper motions to find
membership probabilities ($P_M$) for the brighter members of the cluster.
Subsequent $\ubv$ studies \citep{goetz,sj,cn} found similar colour-magnitude
diagrams: earlier-type stars (to about A0) along the zero-age main
sequence (ZAMS), with later stars lying above it, although \citet{hiltner}
classified the central B-stars of the cluster as mainly Be-type, and
placed them above the ZAMS. \citet{scb} identified
a ZAMS down to about 3~M$_\odot$, and \citet{dam} suggest that all stars
earlier than G-type and with $V<13$ should be considered probable cluster
members. Many stars in the field were excluded from membership by
\citet{vaj}, but \citeauthor{dam} considered only a
$17^\prime\times 17^\prime$ field in the centre of the cluster, compared
to the half-square-degree field of \citet{vaj}. New proper-motion studies
of the cluster used photographic plates of NGC\,6530 taken at Shanghai
Astronomical Observatory in 1912 to obtain a baseline of 87 years
\citep{zhao,wen,chen}, yielding a very clear separation between cluster
and field stars. Based on this selection, \citet{chen} measured core and
half-number radii of $4.3^\prime\pm 0.9^\prime$ ($1.6\pm0.3$~pc)
and $21^\prime$ ($7.6$~pc), respectively; they estimated the cluster
radius (at which the cluster population disappears into the field star
population) to be around $20^\prime$. The cluster density profile is
consistent with either a King or $1/r$ model.

\begin{table}[tbh]
\caption{Optical/IR studies of the stellar content of NGC\,6530}
\label{tab-clusterstudies}
\smallskip
\begin{center}
{\small
\begin{tabular}{l@{\hskip8pt}c@{\hskip8pt}c@{\hskip8pt}c@{\hskip8pt}c@{\hskip8pt}c@{\hskip8pt}l}
\tableline
\noalign{\smallskip}
Survey & Ref\tablenotemark{a} &Area &Stars& Selection & Photometry &
Other Data\\
\noalign{\smallskip}
\tableline
\noalign{\smallskip}
Walker  & 1  & NGC\,6530  & $118$ & $V<16$
& $\ubv$ & \\
HMN     & 2 & NGC\,6530 & $25$ & $V\la 11$ & --- & Spectroscopy \\
VAJ     & 3  & NGC\,6530  & $363$ & $V\la 13$
& Photographic & Proper Motion \\
Kilambi & 4  & NGC\,6530  & $319$ & $V\la 15$
& $\ubv$\tablenotemark{b} & \\
SJ      & 5  & NGC\,6530  &  $88$ & $P_M>0.5$
& $\ubv$ & \\
CN      & 6  & NGC\,6530  & $110$ & $P_M>0.01$\tablenotemark{c}
& $\ubv$\tablenotemark{d} & \\
MRV     & 7  & NGC\,6530  &  $81$ & $V\la 12$
& --- & Polarization \\
VdA     & 8  & NGC\,6530  & $132$ & $P_M>0.1$
& {\it WULBVRIJHK}\tablenotemark{e} & Spectroscopy \\
SCB     & 9  & NGC\,6530  & $887$ & $V<17$
& $\ubvri$, H$\alpha$ &  \\

\multirow{2}{*}{KSSB}   & \multirow{2}{*}{10}  & \multirow{2}{*}{NGC\,6530}  & \multirow{2}{*}{$45$}  & \multirow{2}{*}{$P_M>0.5$} & $\ubvri$\tablenotemark{f}, $JHK_{\rm s}$\tablenotemark{g}, & \multirow{2}{*}{Spectroscopy} \\
 &  &  &  &  &  mid-IR\tablenotemark{h} & \\

Damiani  & 11  & NGC\,6530  & $731$ & X-ray/NIR
& $BVI$\tablenotemark{f}, $JHK_{\rm s}$\tablenotemark{g} & X-ray \\
PDMS    & 12 & NGC\,6530  & $828$ & $V<22$
& $BVI$, $JHK_{\rm s}$\tablenotemark{g} & X-ray \\
ABMMR   & 13 & Hourglass & $763$ & $K_s\la 16$
& $JHK_s$ & \\
ZCW     & 14--16 & NGC\,6530  & $364$ & $B\la 13$ & Photographic
& Proper Motion \\
ZW      & 17,18 & NGC\,6530 & $30$ & HR Diag. & $BV$ & Time-domain \\
ABM     & 19 & NGC\,6530  & $46$  &
& --- & Spectroscopy \\
PDMP    & 20 & NGC\,6530  & $332$ &
&     & Spectroscopy \\
Mayne   & 21,22 & NGC\,6530 & -- & X-ray, H$\alpha$ & V, I & \\
\noalign{\smallskip}
\tableline
\noalign{\smallskip}
\multicolumn{7}{l}{\parbox{0.95\textwidth}{\footnotesize
    $^a$ References: (1) \citet{walker}; (2) \citet{hiltner};
  (3) \citet{vaj}; (4) \citet{kilambi}; (5) \citet{sj};
  (6) \citet{cn}; (7) \citet{mccall}; (8) \citet{vda};
  (9) \citet{scb}; (10) \citet{kumar}; (11) \citet{dam};
  (12) \citet{pris}; (13) \citet{arias05}; (14) \citet{zhao};
  (15) \citet{wen}; (16) \citet{chen}; (17) \citet{zwintz};
  (18) \citet{daveg}; (19) \citet{arias07}; (20) \citet{pris07};
  (21) \citet{nathan1}; (22) \citet{nathan2} }}\\
\multicolumn{7}{l}{\parbox{0.95\textwidth}{\footnotesize
    $^b$~photographic;
    $^c$~4 sources have no known $P_M$
    $^d$~Non-standard photometric system
    $^e$~Walraven, Johnson-Cousins and near-IR systems
    $^f$~from literature;
    $^g$~from 2MASS;
    $^h$~from ISO, MSX, IRAS}}\\
\end{tabular}
}
\end{center}
\end{table}

Most stars near the cluster centre with $V<11$ have known spectral types
\citep[e.g. ][]{hiltner},
but few with $V>11$ \citep{scb}; \citet{dam} found only 68 spectral types
in the literature over their 300 square arcmin X-ray field of view around
the cluster centre, most with $V<13$. Thus there are few intermediate- to
late-type stars with known spectral types.
\citet{vda} published photometry from the near-UV to near-IR for all 132
stars with $P_M>0.1$, and optical spectroscopy for some of them. Their
Table~1 is a comprehensive summary of stellar data from the literature
for this sample. They generally used photometric measurements to estimate
the spectral type, unless the spectroscopic classification was very
different.

\subsection{The Pre-Main Sequence Population}

\label{sec-ngc6530-pms}

\citet{walker} suggested that stars lying significantly above
the ZAMS in NGC 6530 were still contracting onto the main sequence,
but \citet{the} disagreed, arguing that only bright stars showed any sign
of clustering \citep[following][]{wallenquist}, and that data from nearby
non-cluster fields produced colour-magnitude diagrams very similar to
Walker's, but without the early-type ZAMS. The problem with the
clustering argument is that, towards the cluster centre and {\sc Hii}
region, only the brightest stars were readily visible on photographic
plates, giving rise to selection effects. \citet{cn} obtained $\ubv$
photometry for stars with high probability of
cluster membership \citep{vaj}, which showed probable members in the
earlier-type ZAMS and probable non-members in the later-type population,
consistent with Th\'e's scepticism. Spectroscopic classifications
of 11 of these stars \citep{walker61} ruled out their being background
giants, but did not exclude their being foreground stars.

\citet{kilambi} added stars to the Walker sample: Early-type stars
were added by membership probability \citep{vaj}, and later-type stars
by looking for a population whose apparent magnitudes were linearly
related to their absolute magnitude (and hence, presumably, at the
same distance). The colour-magnitude diagram derived from this
expanded list was similar to Walker's, suggesting that there was indeed
a PMS population. \citet{sj} carried out a very similar
study ($\ubv$ photometry of 88 stars with high membership probability)
and again found an early-type ZAMS and late-type PMS.

\citet{dam} found that optical sources do not cluster significantly
(apart from OB stars), whereas X-ray sources in NGC\,6530 cluster rather
strongly. In addition, the age spread of PMS stars smears them out over
the H-R diagram. They conclude that the optically-visible cluster
stars are diluted so much by the field star population, both on the sky
and in a colour-magnitude diagram, that they are unlikely to show up.
In other words, although there is a PMS population, it probably
was not detected by the early optical surveys.

\subsubsection{H$\alpha$ Observations of the PMS Population}

A popular method of isolating the PMS population in the cluster is to
select stars by their H$\alpha$ emission. Earlier, shallower, studies
found brighter objects, mainly HAeBe stars: \citet{herbig} found 19
H$\alpha$ stars in the cluster (LkH$\alpha$~102--119 and the OB star
MWC\,280/HD 164906). An
objective-prism survey \citep{velghe}, covering a $5\deg$~field (including
both M\,8 and M\,20), revealed 66 H$\alpha$ emission objects, mainly early-type
stars. \citet{vda} found 5 stars with intrinsic H$\alpha$ emission and
near-IR excess (probably HAeBe) in their early-type spectroscopic sample.

\citet{scb} conducted a deeper search, comparing narrow-band H$\alpha$
to a broad-band $R$ photometry, and finding 37 PMS stars and 9
candidates from ground-based observations (of which 8 are members of the
Herbig list) and a further 21 much fainter PMS stars (with another 8
candidates) from archival HST images. \citet{arias07} confirmed several
of these stars to be PMS by spectroscopy, and estimated their masses to
be in the range 0.8--2.5~M$_\odot$, in agreement with their position on
an H-R diagram.

\begin{figure}
\plotfiddle{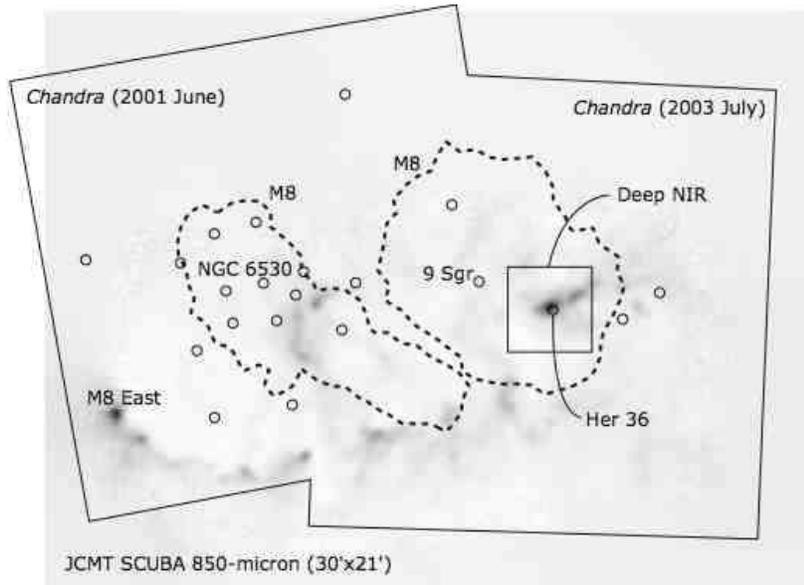}{3.in}{0}{55.}{55.}{-160}{0}
\caption{Submillimetre continuum map of M\,8 (grayscale), overlaid with:
The combined {\it Chandra} field of view (large squares), positions of
bright stars (open circles), extent of bright nebular continuum emission
(broken outlines), and the field of the deep near-IR imaging in
\citet{gagne}. North is up and East to the left; each {\it Chandra} field
is 17\arcmin\ on each side.}
\label{fig-submm-chandra}
\end{figure}

\subsubsection{X-ray Observations of the PMS Population}

\citet{rauw} detected a total of 220 X-ray sources with XMM (119 with high
confidence), primarily in and around NGC\,6530, including 9\,Sgr and
Herschel~36. Nearly all of these high-confidence XMM sources are
associated with candidate cluster members in the optical catalog of
\citet{scb}; few of the XMM-detected stars have strong H$\alpha$ emission,
suggesting that most are low-mass, weak-lined T~Tauri stars. The short
exposure time and the relatively large XMM PSF meant that only a small
fraction (less than 10\%) of the cluster members were detected.

Deeper wide-field images have been obtained with {\it Chandra}-ACIS
(see Fig.~\ref{fig-submm-chandra}): NGC\,6530 was observed in
2001 \citep{dam} and another field centred on the Hourglass in
2003 \citep{gagne}. \citet{dam} found 884 X-ray sources in their
17$\arcmin\times$17$\arcmin$ NGC\,6530 field: only 220 of these have
optical counterparts in the SCB survey \citep{scb}, but the great
majority (731) have counterparts in the 2MASS point source catalog.
With 30--50 of these expected to be spurious, at least 90\% are T~Tauri
stars in the cluster, many of them showing IR excess and some with UV
excess \citep{dam,pris,dam06}. The resulting optical H-R diagrams of X-ray
selected T~Tauri stars suggest an age gradient from the northwest to the
south in NGC\,6530, with the youngest stars located near the southern edge
of the {\sc Hii} region. This result is supported by optical spectroscopy
of PMS stars \citep{arias07}, which shows younger stars along the southern
rim and near the Hourglass. \citet{dam} also noted that {\it Chandra}
sources at the southern edge of the cluster near the ionization front were
harder and more time-variable than those near the center of NGC\,6530
inside the cavity. These results suggest both sequential star formation and
increased flare activity among newly-formed stars.

This sample increases the number of probable cluster members known,
and massively increases the number of low-mass (and hence PMS) cluster
members known, by more than an order of magnitude. \citet{pris} obtained
deeper optical data than that of \citet{scb} to compare with the X-ray
sources: The X-ray-selected subsample of their optical sample is strongly
concentrated spatially to the centre of the cluster and, in
colour-magnitude space, to an isochrone. \citeauthor{pris} used the
Orion Nebula as a well-studied surrogate to argue for a completeness
ranging from below half (for stars below 0.25~$M_\odot$) to about unity
for stars of solar mass and above.

{\it Chandra} can detect only a fraction of the YSOs because of exposure
time limitations and intrinsic variations in X-ray luminosity. The total
(detected and undetected) population of NGC\,6530 and the Hourglass Nebula
may be estimated by comparison to the COUP sample of YSOs in the Orion
Nebula \citep{getman}: For stars with relatively low obscuration, the COUP
data are essentially complete to the hydrogen-burning limit. The complete
Orion X-ray luminosity function (XLF) may then be compared to the (censored)
XLF in NGC\,6530 and the Hourglass Nebula, using limits on the distance and
extinction to estimate minimum and maximum numbers of cluster stars: If
$d\sim 1.8$~kpc, then the limiting X-ray luminosities, given as
log($L_X/$erg~s$^{-1}$), near NGC\,6530 and Herschel\,36 are 29.9 and 29.4,
assuming $A_V\approx 4$ (probably too high for NGC\,6530, but appropriate
for the Hourglass). In this case, the Hourglass would contain approximately
400 cluster members and NGC\,6530 nearly 3000. If $d\sim 1.3$~kpc and
assuming low to medium obscuration, then 70\% of cluster members should
have been detected by {\em Chandra}. The low X-ray detection fraction of
IR sources in the Hourglass is probably due to high obscuration, but could
also be explained by the distance being $>1.3$~kpc. Our adopted distance
of $\sim 1.3$~kpc suggests fewer NGC\,6530 cluster members: $\sim 2000$ in
the field of the 2001 {\it Chandra} observation.

\citet{dam} found that a significant fraction of strong H$\alpha$ emission
stars are not detected in X-rays, which suggests that the different methods
of selecting PMS stars are complementary. While H$\alpha$ is expected to
trace Classical T~Tauri stars preferentially, they suggested that X-ray
flaring sources are an even better tracer of very young pre-main-sequence
populations, and that deeply embedded (and hence very young) stars might be
detected as hard X-ray sources. The two hard, flaring X-ray sources they
found are both in the southeast of the cluster, near M8\,E (a major locus
of star formation).

\vspace{-2ex}
\subsubsection{Optical/near-IR Observations of the PMS Population}

Since X-ray detection does not provide a complete census of the
PMS population, \citet{dam06} combined their previous X-ray and optical
data \citep{dam,pris} with 2MASS to search for more young stars.
Constructing reddening-free optical-IR colour indices (by analogy with
$Q$\footnote{\citet{johnsonmorgan} defined $Q$ as a reddening-free linear
combination of $U-B$ and $B-V$, which acts as a spectral-type diagnostic;
\citet{dam06} define $Q_{VIIJ}$ as a similar combination of $V-I$ and $I-J$}),
they find 196 stars, of which 120 were not previously
detected in X-rays. The majority of these seem to be young stars with
circumstellar disks, but a small subset have rather different colours:
These `strong $Q_{VIIJ}$' objects, are interpreted as candidate Class~I
objects. Their very low X-ray detection rate would then be due to
their still being surrounded by an envelope, and their optical
brightness would be due mainly to reflection nebulosity.

These candidate Class~I objects are mainly found in the northwest of the
\citeauthor{dam} {\it Chandra} field, i.e.\ just northeast of the Hourglass.
This is difficult to reconcile with earlier results \citep{dam,pris}
which show older stars in the north and the youngest sources in the
southeast, potentially associated with the molecular gas there.
\citet{dam06} suggest that this northern region of the cluster might
be undergoing a prolonged process of star formation with
star formation rates and long disk lifetimes (due to a lack of nearby OB
stars), in contrast to the faster star formation elsewhere in the cluster.
The claim of long disk lifetime needs to be examined carefully, since
the position of these sources NE of the Hourglass places them fairly close
to 9\,Sgr. It would also be useful to check whether these sources might
be associated with the Hourglass Nebula Cluster.

\subsubsection{Optical Spectroscopy of PMS Stars}

Until recently, spectroscopy of individual cluster members was restricted
to the brightest stars, generally already on the MS. In their sample
of 45 early-type stars, \citet{kumar} found only 4 probable PMS stars ---
One classical Be and 3 HAeBe (including LkH$\alpha$\,112 and MWC\,280).
Recently, larger
telescopes and multi-object spectrographs have enabled spectroscopic
studies of lower-mass PMS stars. \citet{arias07} selected 46 target stars
by their H$\alpha$ emission \citep[from][]{scb}, near-IR colours or
proximity to optical nebular features. They classified all but 7 of them
as PMS stars: 2 HAeBe, 3 PMS Ge, 27 CTTS and 7 WTTS (the preponderance of
CTTS presumably being due to the H$\alpha$ selection criterion).
\citet{pris07} obtained lithium spectra of 332 PMS candidates, selected
to lie in the same part of the $V$,$V-I$ diagram as the bulk of detected
X-ray sources \citep{dam}, and H$\alpha$ spectra of 115 of them. Using
X-ray detection, radial velocity, and lithium equivalent-width criteria,
they found 237 stars to be certain cluster members (of which 53 are
binaries), and another 10 possible members. The 71 certain members for
which they have H$\alpha$ spectra comprise 31 CTTS, 9 possible CTTS,
and 31 WTTS.

\subsubsection{Time-domain Observations of PMS Stars}

\citet{zwintz} monitored two fields towards NGC\,6530 to look for PMS
pulsating stars (similar to the post-MS $\delta$~Scuti stars). Their
fields cover the core of the cluster (extending out to M8\,E and
including the SE rim) and the area NE of the Hourglass. Of the 30 stars
lying in the classical instability strip of the HR diagram, 6 were
confirmed to be pulsating, with one further candidate. The oscillation
spectra of 5 of the confirmed pulsating PMS stars were then modelled
\citep{daveg}: The best-fit models were somewhat redder than the stellar
colours taken from the literature, but the luminosities were very similar
to those expected, assuming the distance to the cluster to be 1.8~kpc.

\subsubsection{Individual PMS Stars}


\citet{pereira} identified LkH$\alpha$~117 with SS73~125 \citep{sanduleak},
and, based on optical spectroscopy, classified it as a mid-K-type T~Tauri star.
MWC\,280/HD\,164906 (see Sect.~\ref{sec-overview-cluster}) seems to be a
very high-mass HAeBe star \citep{herbig, kumar}, with uncertain membership.
LkH$\alpha$\,108, 112 and
115
 are Herbig Be stars with high cluster
membership probabilities and IR excesses \citep{boesono, vda, scb, kumar},
while NGC6530-VAJ\,45 and NGC6530-VAJ\,151 are HAeBe candidates \citep{vda}.
LkH$\alpha$\,113 had the strongest H$\alpha$ emission in the cluster at the
time of the \citet{scb} survey, but not much else is known about it.
LkH$\alpha$\,109 (SV~Sgr) is a variable
H$\alpha$ star whose proper motion is inconsistent with that of the
cluster, and hence is either a foreground star or a cluster member ejected
by dynamical processes \citep{scb}, and Walker~29 (V5100~Sgr) is a classical
Be star \citep{kumar}, with only a 20\% probability of membership.

\citet{arias07} found 37 new PMS stars by spectroscopy. Stars of particular
interest are: ABM\,22 (SCB\,422), which displays Herbig-Haro emission; and
ABM\,21, 27 \& 29 (SCB\,418, 486 \& 495), which appear to be located in
knots of highly-excited gas. Among their PMS stars are LkH$\alpha$\,108,
111 \& 115
(Table~\ref{tab-pms_stars}); data for the rest may be found
in the original publication.

\begin{table}[tbp]
\caption{Selected PMS Stars and Candidates in NGC\,6530}
\label{tab-pms_stars}
\smallskip
\begin{center}
{\small
\begin{tabular}{lccccc}
\tableline
\noalign{\smallskip}
Primary & R.A. & Dec. & Classification & Other & Membership \\
Name & (J2000.0) & (J2000.0) &  & Name & Prob. \\
\noalign{\smallskip}
\tableline
\noalign{\smallskip}
LkH$\alpha$\,103 & 18:02:51.1 & --24:19:23\tablenotemark{b} &  & & \\
LkH$\alpha$\,102 & 18:02:52.5 & --24:18:44\tablenotemark{a} &  & & \\
HD\,314900       & 18:02:53.3 & --24:20:17\tablenotemark{c} &
HAeBe (B5Ve)\tablenotemark{d} &
Walker\,5 & 2\%\tablenotemark{e}, 90\%\tablenotemark{f} \\
LkH$\alpha$\,104 & 18:02:54.3 & --24:20:56\tablenotemark{a} &  & & \\
LkH$\alpha$\,106 & 18:03:40.3 & --24:23:20\tablenotemark{g} &  & & \\
LkH$\alpha$\,108 & 18:03:50.8 & --24:21:11\tablenotemark{f} &
HAeBe (B6Ve)\tablenotemark{h} &  &
77\%\tablenotemark{e}, 99\%\tablenotemark{f} \\
LkH$\alpha$\,109 & 18:03:57.7 & --24:25:33\tablenotemark{i} &  & SV Sgr &
0\tablenotemark{e} \\
Walker\,13       & 18:04:00.2 & --24:15:03\tablenotemark{p} &
pulsating\tablenotemark{q} & & \\
Walker\,28       & 18:04:09.9 & --24:12:21\tablenotemark{p} &
pulsating?\tablenotemark{q} & & 99\%\tablenotemark{f}\\
Walker\,29       & 18:04:11.2 & --24:24:48\tablenotemark{g} &
B2e\tablenotemark{d} & V5100~Sgr & 20\%\tablenotemark{e} \\
LkH$\alpha$\,110 & 18:04:11.4 & --24:27:16\tablenotemark{g} &  & & \\
Walker\,38       & 18:04:14.0 & --24:13:28\tablenotemark{p} &
pulsating\tablenotemark{q} & & 68\%\tablenotemark{e}, 99\%\tablenotemark{f}\\
LkH$\alpha$\,111 & 18:04:17.5 & --24:19:09\tablenotemark{a} &
CTTS (K5)\tablenotemark{h} &  & \\
Walker\,53       & 18:04:20.7 & --24:24:56\tablenotemark{p} &
pulsating\tablenotemark{q} & & 78\%\tablenotemark{e}, 99\%\tablenotemark{f}\\
Walker\,57       & 18:04:21.8 & --24:15:47\tablenotemark{p} &
pulsating\tablenotemark{q} & & 0\tablenotemark{e} \\
LkH$\alpha$\,112 & 18:04:22.8 & --24:22:10\tablenotemark{f} &
HAeBe (B2Ve)\tablenotemark{d} &
Walker\,58 & 81\%\tablenotemark{e}, 98\%\tablenotemark{f}\\
HD\,164906       & 18:04:25.8 & --24:23:08\tablenotemark{j} &
HAeBe (B0Ve)\tablenotemark{k} & MWC\,280 & 94\%\tablenotemark{f} \\
LkH$\alpha$\,113 & 18:04:26.1 & --24:22:45\tablenotemark{g} &  & & \\
Walker\,78       & 18:04:30.8 & --24:23:42\tablenotemark{p} &
pulsating\tablenotemark{q} & & \\
LkH$\alpha$\,114 & 18:04:33.2 & --24:27:18\tablenotemark{g} &  & & \\
LkH$\alpha$\,107 & 18:04:36.5 & --24:19:14\tablenotemark{b} &  & & \\
Walker\,159      & 18:04:42.3 & --24:18:04\tablenotemark{p} &
pulsating\tablenotemark{q} & & \\
LkH$\alpha$\,115 & 18:04:50.6 & --24:25:42\tablenotemark{f} &
HAeBe (B2Ve)\tablenotemark{h} & &
79\%\tablenotemark{e}, 98\%\tablenotemark{f} \\
LkH$\alpha$\,116 & 18:04:58.6 & --24:24:36\tablenotemark{l} &  & & \\
LkH$\alpha$\,117 & 18:05:39.0 & --24:30:40\tablenotemark{m} &
TTS (K)\tablenotemark{n} & & \\
LkH$\alpha$\,118 & 18:05:49.7 & --24:15:21\tablenotemark{o} &  & &
98\%\tablenotemark{f}\\
LkH$\alpha$\,119 & 18:05:56.5 & --24:16:00\tablenotemark{f} &  & &
0\tablenotemark{e}, 97\%\tablenotemark{f} \\
\noalign{\smallskip}
\tableline
\noalign{\smallskip}
\multicolumn{6}{l}{$^a$ from \citet{ducourant};~~$^b$ position from
2MASS catalogue} \\
\multicolumn{6}{l}{$^c$ from USNO catalogue (UCAC2);~~$^d$ from
\citet{kumar}} \\
\multicolumn{6}{l}{$^e$ from \citet{vaj};~~$^f$ from \citet{zhao},
\citet{chen}} \\
\multicolumn{6}{l}{$^g$ from \citet{pris};~~$^h$ from \citet{arias07};
~~$^i$ from HBC} \\
\multicolumn{6}{l}{$^j$ from PPM \citep{ppm};~~$^k$ see
section~\ref{sec-overview-cluster}} \\
\multicolumn{6}{l}{$^l$ from WCS of DSS image;~~$^m$ from \citet{sanduleak}}\\
\multicolumn{6}{l}{$^n$ from \citet{pereira};~~$^o$ from \citet{teixeira}} \\
\multicolumn{6}{l}{$^p$ from \citet{scb};~~$^q$ from \citet{zwintz}} \\
\end{tabular}
}
\end{center}
\end{table}

\subsection{Extinction and Reddening towards NGC\,6530}
\label{sec-ext-red}

\subsubsection{Extinction}

Since NGC\,6530 has largely been studied by optical \ubv\ photometry,
estimates of the optical extinction towards it are usually obtained
by deriving $E(B-V)$, the colour excess, and multiplying by
$R=A_V/E(B-V)$, the ratio of total to selective extinction.
Table~\ref{tab-clusterparams} shows the estimates of colour excess
towards the cluster from the various photometric surveys, generally around
0.3. Most of these studies estimate the intrinsic colour from broadband
photometry, which may be very inaccurate, although \citet{scb} used
30 early-type stars with known spectral types. \citet{vda} used
their photometry (checked against optical spectroscopy) to fit a model
SED: $E(B-V)$, $R$, and distance were among the fitted parameters.
This approach has the advantage of using more than one or two data
points in wavelength to determine stellar parameters.

The extinction varies from star to star: \citet{vaj} found 3 stars with
$E(B-V)=0.5$, significantly higher than the rest of their sample, and
\citet{sj} found some evidence for systematic extinction gradients
(0.25--0.48~mag) over the cluster field, though most of their more
extreme values are based on very few stars. \citet{vda} found a foreground
extinction of 0.3~mag, along with large, variable (and presumably
circumstellar) extinction towards individual stars. \citet{scb} derived an
average $E(B-V)$ of 0.35 \citep[also adopted by][]{pris}, although the stars
in their sample span the range from 0.25 to 0.5. They dereddened later-type
stars in their study by assuming them to have the same extinction as the
nearest early-type star with an extinction estimate; if \citeauthor{vda} are
correct that the variable extinction is circumstellar, this approach is
unlikely to be accurate. \citet{nathan2} used a modified
Q-method\footnote{The modified Q-method uses updated reddening vectors and
isochrones in colour-colour space, and considers the effects of binarity
and metallicity} to fit individual extinctions to the stars in their sample,
and found an average $E(B-V)$ of 0.33, agreeing well with the extinction of
0.32 that they found by isochrone-fitting.

\citet{mccall} used optical polarisation towards bright cluster members
to analyse the extinction: Observations of two foreground K giants give
a foreground extinction of $E(B-V)=0.17$~mag. The lack of cluster stars
with $E(B-V)<0.27$ led them to postulate the existence of an additional
sheet of obscuring material in front of the cluster, with $E(B-V)=0.1$;
the value of $R$ for this sheet is unclear. \citet{the} estimated
$A_V\sim 2.2$ for the background cloud to NGC\,6530, but this is an
underestimate, at least in the central regions: \citet{pris} and
\citet{arias05} found background stars in their near-IR samples with $A_V$
of 10 to 20, and the near-IR extinction map of \citet{dam06} shows
consistently higher values.

The extinction $A_V$ towards the proplyd G\,5.97--1.17 is about
5~mag \citep{stecklum}, consistent with most determinations of the
extinction towards Herschel~36 (assuming $R=5.6$). \citet{arias05}
found a foreground extinction of $A_K=0.36$~mag ($A_V=3.2$~mag, for
normal reddening) towards the Hourglass region. By selecting probable
background stars and estimating their extinction, they also mapped the
extinction, showing a strong congruence with the molecular data.

\subsubsection{Reddening}

Reddening is usually measured by $R$, whose canonical value of
about 3.1 appears to be valid over most of our Galaxy. However, larger
values of $R$ have been found in and around star-forming regions. A
larger value implies slightly larger dust grains, which do not block
blue light quite as efficiently. Anomalously high
values of $R$ in M\,8 may be explained by selective evaporation of small
grains by the radiation from hot stars, or by grain growth in
circumstellar environments. The former could explain abnormal reddening
throughout NGC\,6530 (if there is any), and the latter might produce
the star-to-star variations seen by \citeauthor{vda}

\citet{walker} suggested that the extinction law might be abnormal,
but later attributed the effect to photometric error \citep{walker61}.
\citet{cn} found no evidence of abnormal reddening, and \citet{nc} used
photometric observations from $U$ to $I$ to show that 4 OB stars in
NGC\,6530 had normal extinction ($R=3.1$), whereas similar stars in other
{\sc Hii} regions had larger values of $R$. Although their photometric
system is non-standard \citep{taylor}, this result is probably still valid.
Based on multiwavelength data towards the double O-star HD~165052,
\citet{arias02} found reddening consistent with a standard reddening law,
though their result assumes a distance of 1.8~kpc, which probably too
high (see sec.~\ref{sec-distance-cluster}); Further observations
towards 6 other OB stars \citep{arias05} yield a range of reddening
estimates, from 3.3 to 5.4 (for Herschel~36). UV observations show
uniformly low extinction towards early-type stars, but normal extinction
in optical wavebands \citep{bv, torres}, implying that the small grains
are depleted; the UV extinction towards early-type stars is quite
variable, with some evidence of systematic variation \citep{bbv}.
Removing a foreground extinction (assumed to have normal
$R$ of 3.2), \citet{mccall} used the cluster method to fit
$R=4.64\pm 0.27$ to the remaining extinction, inconsistent with the
standard extinction laws or, indeed, with the anomalous extinction found
around Herschel~36. \citet{kumar} adopted a similar approach, and derived
a similar foreground-subtracted reddening ($4.5\pm 0.1$).
\citet{vda} found that the majority of their fitted SEDs were consistent
with $R=3.1$; the exceptions ($R>3.2$) also have high extinction, so they
attribute anomalous reddening to circumstellar material. \citet{scb} found
a large range of reddening for their sample of 30 stars, with some sign of
correlation between reddening and extinction, which would tend to support
the circumstellar hypothesis of \citeauthor{vda}. For their recent
photometric studies, \citet{pris} assumed standard reddening, while
\citet{arias05} used a standard reddening law in the near-IR
\citep{rieke} to avoid the non-standard optical reddening.

\subsection{The Age of NGC\,6530}
\label{sec-age}

The usual way to estimate the age of a very young cluster like NGC\,6530
(whose H-R diagram has no giant branch) is to isolate the PMS stars and to
compare them to theoretical isochrones. The earliest attempts to do
this \citep{walker, vaj, kilambi, sj} may not have included any PMS stars in
their samples: The stars that these studies found to the right of the ZAMS
are probably background giants, and the PMS has likely only been isolated by
more recent research. Therefore, these older isochrone-based age estimates
are unlikely to be useful. The H$\alpha$-selected PMS population clusters
around isochrones of order 1~Myr \citep{scb}, and almost none are older than
3~Myr \citep{arias07}. Both \citet{dam} and \citet{pris} placed the X-ray
selected sample on a $(V/V-I)$ diagram: \citet{dam} derived a median age of
0.8~Myr, with a spread of about 4~Myr; \citet{pris} found almost all of their
stars to lie between the 0.3~Myr and 10~Myr isochrones, the distribution
peaking around 2~Myr. The derived ages are almost unaffected by
reddening \citep{dam}, but are strongly model-dependent --- according to some
models, the median age is just 0.1~Myr. The H$\alpha$-selected PMS population
may not be entirely the same as the X-ray selected one: H$\alpha$ is
strongest in classical T~Tauri Stars, whereas X-ray emission mainly comes
from weak-line T~Tauri Stars. However, no systematic difference between the
populations is evident on an H-R diagram \citep[e.g.~][]{arias07}.

\citet{nathan1} and \citet{nathan2} take a different approach: By comparing
the ($V$, $B-V$) colour-magnitude diagrams of multiple young clusters, they
obtain a relative age ladder. According to their results, NGC\,6530 is
indistinguishable from the Orion Nebula Cluster (ONC) in age, older than
IC\,5146, and younger than NGC\,2264. This corresponds to an absolute age of
1--2~Myr.

Although NGC\,6530 has no giant branch, some of the OB stars have started
to evolve off the ZAMS: by fitting 10 OB stars to theoretical tracks,
\citet{bv} estimated an age of $5\pm 2$~Myr. However, this estimate assumes
a distance modulus of 11.5~mag, probably $\sim$1~mag too high. A lower
distance reduces the calculated luminosities, and hence the estimated ages,
though not enough to place the stars on the ZAMS. \citet{walborn73}
classified 9\,Sgr as O4\,V((f)), i.e.~still on the main sequence.
\citet{vda} found at least one star in their sample with an age of 15~Myr
and a high probability of cluster membership, leading them to suggest that
star formation has been going on in NGC\,6530 for a few $10^7$ years.
On dynamical grounds, \citet{vaj} placed a lower limit of 0.6~Myr on
the cluster age. \citet{chen} measured an intrinsic velocity dispersion
of 8~km\,s$^{-1}$; if this is taken to be the expansion velocity, it
suggests a dynamical age of about 1~Myr. This velocity dispersion also
suggests that NGC~6530 could survive as a cluster for some hundreds
of Myr, although external perturbations (which are very likely, given
its position in the Galaxy) could disrupt it earlier \citep{chen}.

It seems that the main burst of star formation in NGC\,6530 occurred about
1--2~Myr ago. However, there may have been significant star formation
activity beforehand; whether it stretched back over the tens of Myr
proposed by \citeauthor{vda} is unclear.

The confirmed O-stars in NGC\,6530 (HD\,165052, 9\,Sgr \& Herschel\,36) all
lie well outside the bright core of the cluster, where the brightest
star is B0. \citet{arias07} found that their younger PMS stars ($<1$~Myr)
were preferentially located towards the southern rim, and in the west,
around the Hourglass. There is evidence of similar age separation in the
X-ray selected PMS population \citep{dam, pris}, with older stars
(a few Myr) concentrated to the northeast, and the youngest stars
(less than 0.5~Myr) in an arc running from the southeast (near M8\,E) to
the southwest of the cluster, stretching towards Herschel~36. This fits
very well with the submillimetre maps that show a broad arc of dense gas
to the south of the cluster centre, and which might be loci of ongoing
star formation \citep{thesispaper}. This interpretation suggests that at
least the youngest part of NGC\,6530 is still embedded within its natal
molecular cloud.

\subsection{The Distance to NGC\,6530}
\label{sec-distance-cluster}

Many authors have estimated the distance to NGC\,6530 by fitting an offset
ZAMS to their colour-magnitude diagrams, where the offset gives the distance
modulus \citep{walker, walker61, vaj, cn, sj}. However, the effects of
extinction must be
removed, which requires the determination of $A_V$, in this case about
1 mag. This process gives considerable room for error, especially in
light of the uncertainty over the amount and nature of extinction
towards NGC\,6530.

\citet{mccall} used published spectral types for early-type stars to
derive the extinction and distance, using the `cluster method', which
attempts to distinguish binaries from single stars. \citet{scb}
also used stars with known spectral types, finding the stars to have
either low ($\sim 10.75$) or high ($\sim 11.25$) distance moduli.
They adopted the higher value on the grounds that the low distance
estimates are probably binaries.
\citet{vda} do not use a distance modulus at all, but instead directly
fit a theoretical SED to their (spectroscopically-checked) multi-band
photometry, with distance as
one of the parameters. They base their distance estimate of 1.8~kpc
on the histogram of fitted distances.

More recent studies have generally found smaller distances: \citet{pris}
fitted a ZAMS at 1.3~kpc to the blue edge of the stellar distribution in
colour-magnitude diagrams. They argue that any field stars significantly
further away than NGC\,6530 will lie behind the molecular cloud, and thus
be highly reddened, moving away from the ZAMS on the colour-magnitude
diagram. Thus the blue edge of the distribution should be defined by the
ZAMS at the distance of NGC\,6530. \citet{arias05} also derive a low
distance (1.3~kpc), based on half a dozen early-type stars near the
Hourglass, whose distance moduli and extinctions were derived by fitting
stellar models to their SEDs. The spread in derived extinctions is large,
probably due to the highly variable interstellar and circumstellar
extinction, as is the spread in distance modulus. All three O stars in the
sample are binaries (including 9\,Sgr and Her~36), and this only seems to
have been taken into account for one of them; the three B stars have
rather higher distance moduli. \citet{dam} also note that their estimate
of the number of foreground field stars is close to that which might be
expected if the cluster were 1.3~kpc away; if it were 1.8~kpc away, the
expected number of foreground stars would be twice as high. \citet{nathan2}
used a modified Q-method (see also Sect.~\ref{sec-ext-red}) to analyse
NGC\,6530, rather than their preferred $\tau^2$ \citep{tau-squared},
because of the large spread in extinction over the different cluster
members. They derived a distance of about 1.3~kpc, with an error range of
$<0.1$~kpc, using photometry from \citet{scb}.

One recent study, however, supports a greater distance: \citet{daveg}
found that their modelling of the oscillation spectra of PMS pulsating
stars gave luminosities consistent with their assumed distance of 1.8~kpc.
{\it Hipparcos} parallax measurements, on the other hand, suggest a
distance of only about 600~pc
\citep[distance modulus $9.01\pm 0.26$,][]{lotkin} towards NGC\,6530.
Parallax measurements are generally considered to be robust only for
nearby objects, and this determination is so different from all other
estimates as to be hard to believe. The measurement is based on only 7
stars; since they are presumably rather bright, it is entirely possible
that they are all foreground stars. \citeauthor{lotkin} were able to fit a
straight line to the relationship between distance moduli determined from
{\it Hipparcos} data and those determined photometrically, but the
NGC\,6530 measurement is inconsistent with that straight line; moreover,
they consider 1~kpc to be the greatest distance at which their technique is
accurate. In the absence of more detailed results, this distance estimate
should not be adopted.

Most recent distance determinations agree on a distance of about 1.3~kpc,
which is used throughout this review. The \citeauthor{scb} estimate of
1.8~kpc is based only on early-type stars for which a spectral type was
known: Essentially, they fitted a ZAMS to the top (the blue end) of the
colour-magnitude diagram. As \citeauthor{pris} point out, the blue end of
the ZAMS is almost vertical on a colour-magnitude diagram, so the distance
modulus (a vertical offset on the diagram) is not very well-constrained.
By contrast, \citeauthor{pris}, by fitting to the blue envelope, are able
to use the redder part of the ZAMS to constrain the distance modulus, at
the cost of the additional assumption that the blue edge of the
colour-magnitude diagram constitutes the cluster, or field stars at the
same distance. \citeauthor{nathan2}, by using the modified Q-method, were
also able to use fainter stars without spectroscopic data, and in fact
used the photometric data from \citet{scb} to derive their nearer distance.
The \citet{vda} estimate (also 1.8~kpc) is similarly dependent on rather
bright stars. However, the differences in stellar samples used for the
various distance determinations may not be the whole story: \citet{scb}
found a distance modulus of 11.2~mag towards Herschel~36, while
\citet{arias05} estimated 10.5~mag. Both determinations used the same
observed magnitude ($V=10.297$), and their assumed absolute magnitudes only
differ by 0.1; the rest of the discrepancy presumably lies in the
assumptions used to deredden the $V$-band data. Herschel~36 is probably an
extreme example, as it is known to have anomalous extinction, but it
illustrates some of the problems that bedevil distance estimation.

Most of these distance estimates are variations on fitting a ZAMS to a
colour-magnitude diagram, but the two recent works which use independent
methods find different distances. As noted above, we do not consider the
parallax measurement \citep{lotkin} to be sufficiently accurate to be
useful, but the asteroseismology of PMS pulsating stars \citep{daveg}
raises some intriguing possibilities. \citeauthor{daveg} assumed a distance
of 1.8~kpc in their analysis, and found well-fitting models. It is unclear
to us whether or not they could have assumed a distance of 1.3~kpc and
still have modelled the cluster stars successfully. The stars whose
oscillations were modelled were selected because they lay on the
instability strip of the HR diagram at an assumed distance of 1.8~kpc
\citep{zwintz}, and it might be useful to search for pulsating stars
outside this region, since the position of the instability strip on a
colour-magnitude diagram could add additional distance constraints.
Further distance estimates could perhaps be obtained from spectroscopy of
known intermediate- and low-mass cluster members (selected by further
proper motion studies, X-ray emission, H$\alpha$ emission, IR excess etc).
If the distance could be shown to be 1.3~kpc rather than 1.8~kpc, this
might further constrain the PMS models used to analyse the pulsating stars
in the cluster.

\begin{table}[tbp]
\caption{Cluster Parameters of NGC\,6530}
\label{tab-clusterparams}
\smallskip
\begin{center}
{\small
\begin{tabular}{lccccc}
\tableline
\noalign{\smallskip}
Survey\tablenotemark{a} & $E(B-V)$/mag & $R$ & Age/Myr & d.m./mag &
Distance/kpc \\
\noalign{\smallskip}
\tableline
\noalign{\smallskip}
Walker  & $0.33-0.37$    & --- & $3$\tablenotemark{b} & $10.7-11.5$  & $1.4-2.0$ \\
VAJ     & $0.35$         & --- & $2$\tablenotemark{b} & $11.0-11.25$ & $1.6-1.8$ \\
Kilambi & $0.35\pm 0.01$ & $3.0$ & $1-3$\tablenotemark{b} & $10.7$ & $1.4$ \\
SJ      & $0.35$         & --- & $>2$\tablenotemark{b} & $11.3\pm 0.1$ & $1.8\pm 0.1$ \\
CN      & $0.36\pm 0.09$ & --- & --- & $11.4$ & $1.9$ \\
MRV     & $0.17$\tablenotemark{c} & $4.6\pm0.3$ & --- & $11.35\pm 0.08$ & $1.86\pm 0.07$ \\
VdA     & $0.3$ & 3.1 & few $\times 10$ & --- & $1.8\pm 0.2$ \\
SCB     & $0.35$ & $>3$\tablenotemark{d} & $1.5, 5$\tablenotemark{e} & $11.25\pm 0.1$ & $1.8\pm 0.1$ \\
KSSB    & --- & $3.9\pm 0.05$\tablenotemark{f} & --- & --- & --- \\
Damiani & --- & --- & $0.8, 4$\tablenotemark{e} & --- & --- \\
PDMS    & --- & --- & $2, 5$\tablenotemark{e} & $10.5$ & $1.3$ \\
ABMMR   & $0.34$, $0.30$\tablenotemark{g} & --- & --- & $10.5$ & $1.3$ \\
Mayne   & $0.33$ & --- & 1--2 & $10.50^{+0.10}_{-0.01}$ &
$1.26^{+0.06}_{-0.01}$ \\
\noalign{\smallskip}
\tableline
\noalign{\smallskip}
\multicolumn{6}{l}{$^a$ References as for
Table~\ref{tab-clusterstudies};~~$^b$ probably unreliable --- see
section~\ref{sec-age}} \\
\multicolumn{6}{l}{$^c$ foreground extinction only;~~$^d$ anomalous,
non-uniform} \\
\multicolumn{6}{l}{$^e$ median age and age spread, respectively} \\
\multicolumn{6}{l}{$^f$ subtracting foreground reddening yields $R=4.5$} \\
\multicolumn{6}{l}{$^g$ foreground extinction towards Hourglass and mean extinction towards early-type } \\
\multicolumn{6}{l}{~~ stars, respectively} \\
\end{tabular}
}
\end{center}
\end{table}

\subsection{Stellar Masses in NGC\,6530}

\citet{pris}, by placing their X-ray-selected sample on a $(V/V-I)$
diagram, estimated the stellar masses. Correcting for the incompleteness
of the X-ray sample, they fitted a mass function with a power law of
index $1.22\pm0.17$. \citet{scb} also fitted power law mass spectra
to their (much smaller) sample, using three different suites of stellar
models, giving indices of 1.2, 1.3 and 1.4. They suggest an overall
index of $1.3\pm 0.1$. Both of these estimates are consistent with the
Salpeter mass function (1.35) and shallower than the mass function of
submillimetre clumps \citep[$1.69\pm0.45$,][]{thesispaper}, although the
stellar and clump mass functions are formally consistent with one another.
PMS mass estimates are strongly affected by reddening \citep{dam}; since
the reddening is poorly characterised in the cluster, this could be a
significant source of error, especially for stars which might be embedded.

\citet{pris} estimated a total stellar mass for NGC\,6530 (down to
$0.4M_\odot$) of 700--930~$M_\odot$. However, a Salpeter IMF with
sufficient amplitude to supply 60 B-stars has a total mass of
$>2000M_\odot$, which suggests that the X-ray sample is very
incomplete. The total stellar mass of NGC\,6530 appears to be of
order $10^3M_\odot$.
\citet{chen} found significant mass segregation in the cluster, with
more massive stars concentrated in the centre.

\section{The Hourglass Nebula}

\label{sec-hg}

\begin{figure}[thb]
\plotone{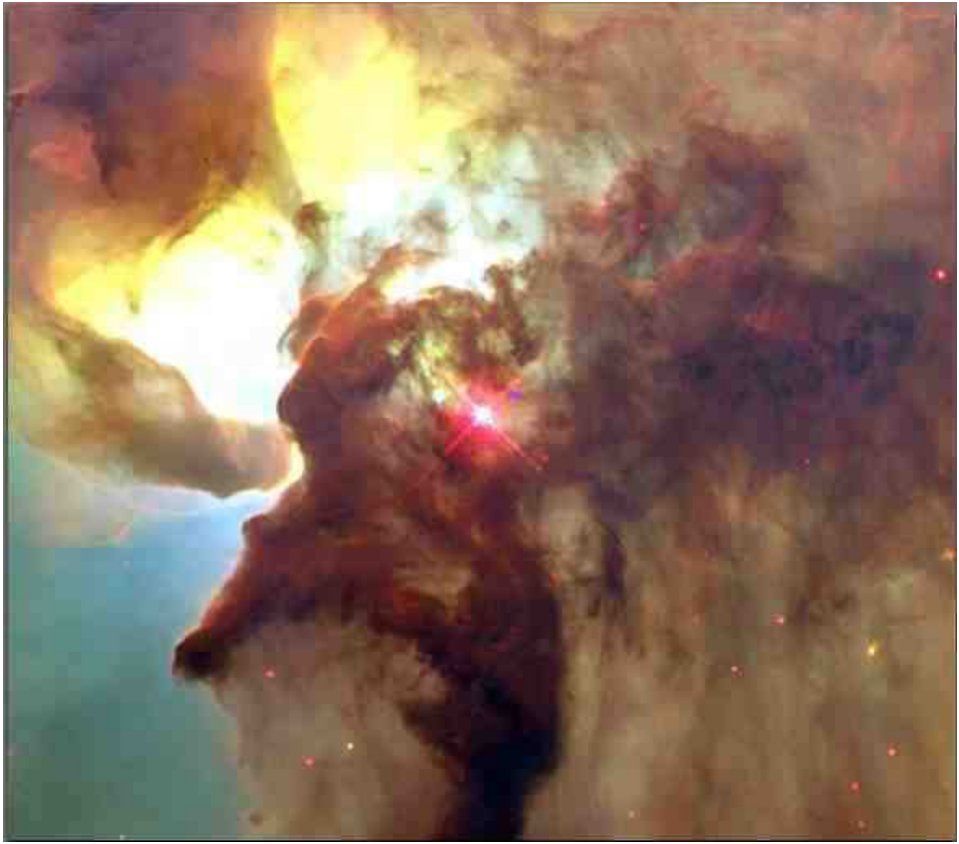}
\caption{HST/WFPC2 image of the Hourglass Nebula and the young massive
star Herschel 36 (in center of image). North is towards the top right,
East towards the top left; FOV is 34\arcsec\ on each side. Courtesy STScI.}
\label{fig-hst-hg}
\end{figure}

The surface brightness peak of the {\sc Hii} region, lying just to the
east of the O7\,V star Herschel\,36, was described by John Herschel as
{\em ``a kind of elongated nucleus, just following a star\dots The proper
nucleus is decidedly not stellar\dots''} \citep{thackeray}. Later
observations showed it to have a narrow-waisted bipolar
appearance \citep{thackeray}, very much like an hourglass (see
Fig.~\ref{fig-hst-hg}). Nearby to the NE, in the outskirts of the
nebular emission, is the star
Cordoba\,12403\footnote{also SCB\,182, CD\,--24$^\circ$\,13810; the
designation Cordoba\,12403 is not used in SIMBAD} \citep[B2\,V,][]{vda}:
\citet{vaj} found this star to have a membership probability
(in NGC\,6530) of 86\%, so it might be associated with the Hourglass, or
might lie in front of the nebula, although still within the NGC\,6530
cluster. \citet{thackeray} found several point-like condensations in the
Hourglass, and mid-IR emission (10--20~$\mu$m) has been found towards
one of these condensations \citep[IRS1,][]{gillett,dyck}, towards
Herschel~36 \citep{woolf,dyck} and Cordoba~12403 \citep{woolf}, and
towards IRS2 \citep{dyck}, located just to the east of the optical
Hourglass, within the obscuring material that defines the eastern edge
of the structure.

Tentative early suggestions that the Hourglass might be a bipolar nebula
around a central star \citep[e.g.][]{allen} are not borne out by detailed
observations. Although the biconical shape of the nebula is maintained
at 2~$\mu$m \citep[][and Fig.~\ref{fig-nir}]{allen}, it disappears at longer
wavelengths, becoming a more rounded blob \citep{woodward}. In this latter
work, \citeauthor{woodward} combine multi-wavelength observations to show
that the Hourglass is an interestingly-shaped window into a compact
{\sc Hii} region lying within a molecular cloud, ionised by Herschel\,36.
The delicate curves and traceries that we see in high-resolution images
(e.g.\ from HST, Fig.~\ref{fig-hst-hg}) indicate the complexity
of the interstellar medium\footnote{It has been suggested (outside the
refereed literature) that such shapes may indicate processes analogous
to a terrestrial tornado, but, in the absence of any theoretical
justification, we consider these claims to be implausible.}.

Although the clear symmetry of the Hourglass disappears at longer
wavelengths, there is a much larger structure with a N--S axis
of symmetry surrounding it --- the Super-Hourglass Structure (SHGS)
of \citet{lada}. This structure is most prominent in e.g.~{\sc [Sii]},
and is probably associated with an ionisation front, most likely
due to 9\,Sgr.

\subsection{Molecular Gas in the Hourglass}

Molecular gas is found in abundance towards the Hourglass Nebula, and the
discovery of a strongly contained {\sc Hii} region inside a clump very
close to Herschel\,36 suggests that the young stars and gas are intimately
associated.

Extinction data \citep{arias05} suggest very high column
density to the north and east of the Hourglass, reflected in a
significant deficit of background field stars. This is consistent
with the molecular line data of \citet{white97}, who
found very high CO brightness temperatures of CO (of order 100~K)
to the N and E of the Hourglass, with weaker emission to the NW,
in M8WC1.

\citet{white97} estimated that the underlying
cloud has a mass of about 31~M$_\odot$, with a column density
$N$(H$_2$)$\sim 10^{23}$~cm$^{-2}$. The large suite of CO
lines observed allowed \citeauthor{white97} to use a
Large Velocity Gradient (LVG) model
to estimate the volume density $n$(H$_2$)$\approx7\times 10^{3}$~cm$^{-3}$.
\citet{thesispaper} used submillimeter continuum data to estimate a mass
of 10--30~M$_\odot$; the column density estimated from this measurement
agrees with the earlier figure, but the volume density is 30 times
higher than the LVG estimate. A gas temperature of 48~K can be derived
from transitions of isotopically-substituted CO
\citep[$^{13}$CO and C$^{18}$O, ][]{thesispaper}.
The discrepancy between this temperature and the
brightness temperature of $^{12}$CO may reflect the effect of
external heating (by the OB stars of the Hourglass Nebula Cluster)
on the molecular cloud, giving
rise to a hot outer layer which dominates the $^{12}$CO
spectrum, but is too thin to make much difference to the more
optically-thin isotopically-substituted transitions.

\subsection{The Stellar Population of the Hourglass Nebula}

\begin{figure}[tbh]
\centering
\includegraphics[draft=False,width=\textwidth]{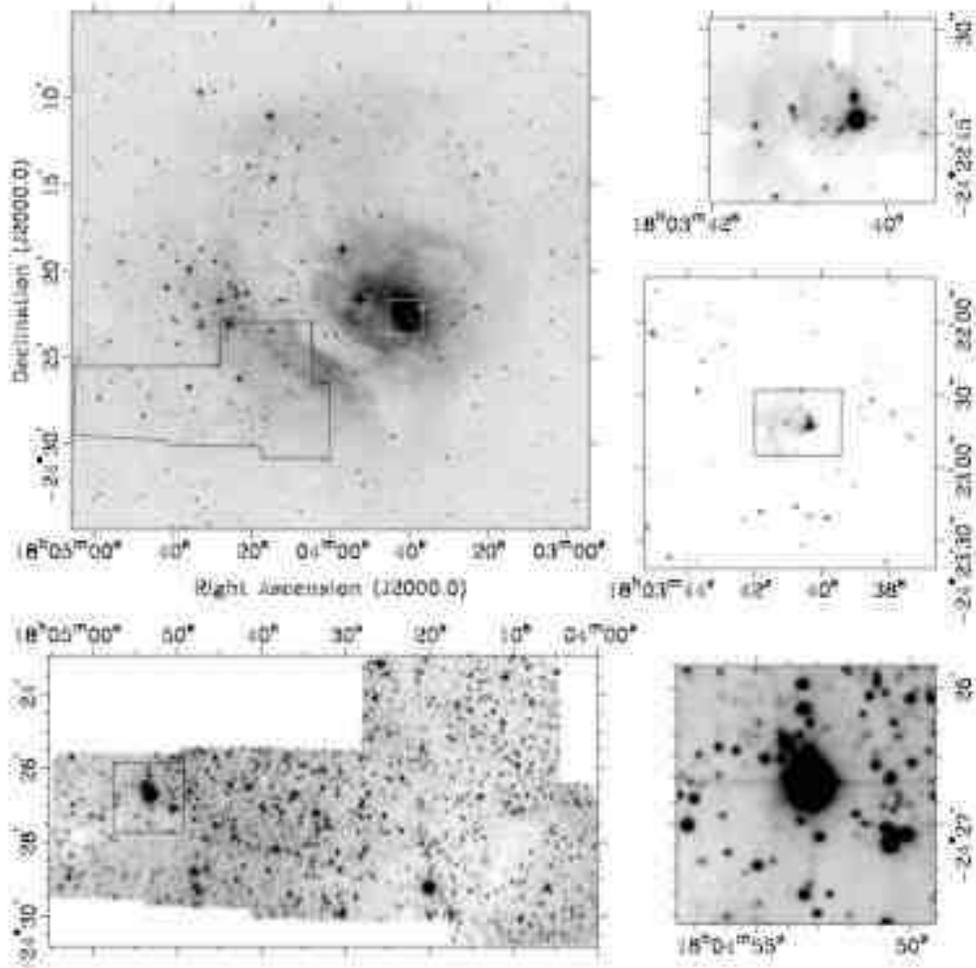}
\caption{POSS image of the Lagoon Nebula with outlines to show the extent
of the near-IR ($K_s$) images of M8\,E and the SE rim ({\it lower left})
and of the Hourglass Nebula ({\it mid right}). The outline around the
Hourglass shows the field covered by the adaptive-optics $K_s$
image ({\it upper right}, from \citeauthor{stecklum}~1998).
M8\,E and its surroundings (outlined) are magnified ({\it lower right}).}
\label{fig-nir}
\end{figure}

Near-IR observations (e.g.~Fig.~\ref{fig-nir}) reveal a large stellar
population around the Hourglass Nebula covering a few arcminutes, with a
concentration (a few arcseconds across) around Herschel~36 itself.

\subsubsection{Herschel 36}

\citet{woolf} found strong extended emission around this young O7 star
throughout the near- and mid-IR spectral range, and this was confirmed at
4~$\mu$m by lunar occultation data \citep{stecklum95}.
Extension along a roughly SE--NW axis has been seen in high-velocity
molecular gas \citep{white97} and more clearly in imaging spectroscopy
of excited H$_2$ \citep{burton}; these results are consistent with outflow
from the star, but by no means conclusive: The broadband near-IR extension
to the SE \citep{stecklum95} is actually a separate object
\citep[][discussed below]{stecklum99,goto}, the submm data lack resolution,
and \citeauthor{burton} points out that the H$_2$ may be excited by
fluorescence rather than shocks. If the H$_2$ emission does come from
an outflow, however, the mechanical luminosity is $\sim 300$~L$_\odot$.

\subsubsection{The Surroundings of Herschel 36}

The first near-IR observations of Herschel~36 \citep{allen,woodwardir}
revealed a small cluster of stars, and later high-resolution
adaptive-optics (AO) observations
\citep[see Fig~\ref{fig-nir}, and][]{goto} resolved more stars.

\citet{woodwardir} found the two pointlike sources at the waist of the
Hourglass (their KS3, 4), along with KS2 which lies between Herschel~36
and the Hourglass, to be consistent with reddened B stars. The AO image,
however, shows KS2 to consist of at least 2 or 3 stars. 3\arcsec\ N of
Herschel~36, KS1/Her\,36B is also resolved into two sources, one visible
at $J$, $H$, and $K_s$, and the other only becoming prominent at $K_s$
and longer wavelengths \citep{stecklum99,goto}.

\citet{stecklum99} found that the SE extension seen in $K^\prime$ is
separated from Herschel~36 by about 0.3\arcsec\ (400~AU) in $L$-band.
A more comprehensive study of this source \citep{goto} found it to be
just 0.25\arcsec away from Herschel~36, extended in broadband emission,
but compact ($<100$\,AU) in H$\alpha$, Brackett-$\gamma$ and radio
continuum emission. \citeauthor{goto} inferred the existence
of an early-type star surrounded by a highly-confined {\sc Hii} region
within a dense clump, and suggested that a B2 star inside a clump with
gas density of order $10^7$\,cm$^{-3}$ would be consistent with their
observations. Further to the SE, \citet{stecklum} showed that the
$K^\prime$ source lying 3\arcsec\ away from Herschel~36, and coincident
with the radio source G5.97--1.17 \citep{woodchurch}, is a young star
(later than B5) surrounded by a circumstellar disc which is being
photoevaporated by the UV flux of Herschel~36, i.e. a proplyd.

Most of these nearby objects lie within a few arcseconds of Herschel~36,
and could account for its supposed unusual reddening, based on aperture
photometry \citep{woodwardir}. Indeed, the closest source lies
within half an arcsecond, and will contaminate all but the
highest-resolution data.

The development of AO imaging, by allowing the Hourglass to be studied at
much higher spatial resolution than before, has reinforced the similarities
between this region and the Trapezium in Orion, the prototypical massive
star-forming region. More than one massive star is found in close proximity,
and nearby star-forming structures are strongly influenced by the UV flux
of the massive stars.

\subsubsection{The Hourglass Nebula Cluster}

Near-IR imaging of the surroundings of the Hourglass Nebula
\citep[e.g.~Fig.~\ref{fig-nir}, ][]{bica,arias05} reveals a much richer
young stellar cluster than previously appreciated. In X-rays, a very
deep {\it Chandra} observation \citep{gagne}, centered on the
Hourglass Nebula, revealed a soft X-ray source at the location of
Herschel~36, surrounded by a tight cluster of harder sources.

\begin{figure}
\plotone{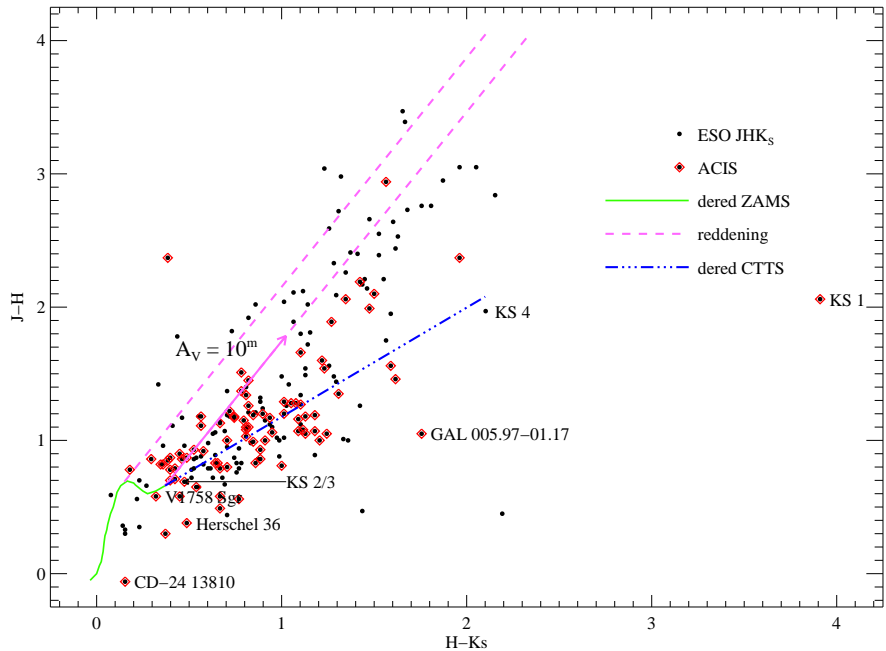}
\caption{Near-IR colour-colour diagram of the Hourglass Nebula
cluster with X-ray detections (diamonds), from \citet{gagne}.}
\label{fig-hgcc}
\end{figure}

\citet{arias05} obtained $JHK_s$ photometry for 945 stars in a
$2\arcmin\times2\arcmin$ region around Herschel~36, of which 102 are
detected with {\it Chandra}. \citet{gagne} used near-IR data from
\citet[][see Fig.~\ref{fig-nir}]{stecklum} to obtain a catalog of
1290 $J$-, $H$- and $K_s$-band sources in a $3\farcm5\times3\arcmin$
region centered on Herschel~36, of which 205 have good $J$, $H$ and $K_s$
photometry, and 128 are detected with {\it Chandra}. The near-IR
colour-colour diagram (Fig.~\ref{fig-hgcc}) shows a cluster containing
numerous reddened T~Tauri stars with large H--K colour excess from disks,
lying to the right of the reddening vector and above the locus of T~Tauri
stars. The diagram also shows a number of YSOs below the T~Tauri locus,
including the proplyd G5.97--1.17 and the bright IR source KS~1. Most of
these sources are detected with {\it Chandra}; the X-ray sources are
tightly clustered  near the 850\,\micron\ emission peak, and may be
Class~I protostars, representing one of the most recent bursts of star
formation in M\,8.

\citet{arias05} used their photometry to disentangle the Hourglass cluster
from the reddened background giant population. Of the 700 or so stars
identified in about 4 square arcminutes around the Hourglass, they found
about 200 potential cluster members, of which about 100 have an infrared
excess (similar to the $\sim$100 stars detected by {\it Chandra}), and
should therefore be considered probable cluster members. Subdividing their
field into 9 areas, they found a large overdensity of potential young stars
in the vicinity of the Hourglass, compared to the outer areas, confirming
the presence of a significant cluster, about 1--2~Myr old. The IR-excess
sources tend to cluster together in a few locations: Around Herschel~36,
around the molecular clump to the NW (M8\,WC1), and on the NE rim of the
Hourglass molecular clump. The subcluster near Herschel~36 extends over
about an arcminute to the south, but hardly at all to the north. This could
reflect the distribution of young stars (suggesting that the Hourglass
itself is the northern extremity of this star-forming region), or it could
be that the southern extension is the only visible part of a more
symmetrical cluster embedded deep in the molecular gas (the southern part
of the cluster lying in the cavity excavated by Herschel~36). The lack of
visible background field stars towards the centre of the clump implies that
the molecular gas is dense enough to extinguish stars even in the near-IR.
High-resolution mid-IR imaging of the Hourglass cluster might find more
deeply embedded members.



\section{M8\,E}

\label{sec-m8e}

The high-mass star-forming region M8\,E was first reported by
\citet{wright} as a 70~$\mu$m continuum source and as a strong CO peak off
the eastern edge of earlier maps \citep{lada}. An 11~$\mu$m source whose
position is consistent with M8\,E appears in the AFCRL catalogue
\citep[CRL\,2059, ][]{afcrl}. The CO data suggest that M8\,E lies within a
large (few arcmin) cloud of molecular gas, with mass of order
$10^4$~M$_\odot$. Bolometric luminosity estimates of
$1.5-2.5\times 10^4$~L$_\odot$ are roughly equivalent to a B0V star
\citep{thronson,mueller}. The available IR to submillimetre continuum data
have been summarised by \citet{mueller}.

Within the molecular gas, there is a small, but quite rich, embedded
cluster: 7 IR sources in  a region of less than a square arcminute have been
catalogued (see Table~6), including a ZAMS B2 star powering a
very small ($0.6\arcsec$ diameter) {\sc Hii} region (M8E-Radio), and, only
7\arcsec\ away, M8E-IR, a massive YSO likely to become a B0 star
\citep{s84,s85,linz}. M8E-Radio is heavily-obscured and, if it is expanding
at the sound speed, has a dynamical age of only 150 years. It has a
cometary morphology, whose leading edge points approximately towards the
bright rim of the M8E clump, although \citet{linz} suggest that it could
have been shaped by the outflow from M8E-IR (discussed below).

Although M8E-Radio is visible in the infrared \citep{s85}, the cluster
luminosity is dominated by M8E-IR up to a wavelength of 24.5~\micron\
\citep{linz}. At longer wavelengths, we lack observations with sufficient
angular resolution to distinguish the fluxes of the two main objects, up to
cm-wave radio, where the {\sc Hii} region is dominant. It is still unclear
how the total luminosity of M8\,E is divided between these two dominant
sources \citep{s85,linz}.

The IR spectral lines observed by ISO \citep{whiteiso} seem to arise from
the {\sc Hii} region and suggest a density of about $10^4$~cm$^{-3}$.
Longward of about 80~\micron, photometric data (measuring the sum of the
luminosities of both components) are well fitted by a model of a spherical
dust envelope with radius of 0.06~pc, temperature 28~K, mass about
80~M$_\odot$, and bolometric luminosity of order $10^4$~L$_\odot$.
Submillimetre-wave continuum mapping \citep{thesispaper,mueller} finds a
source very close to the position of the {\sc Hii} region, with
450~\micron\ and 850~\micron\ fluxes broadly in line with those reported
by \citet{whiteiso}. M8\,E lies at the extreme eastern edge of the SCUBA
maps, where image fidelity is dubious, but is at the centre of the
350~\micron\ map \citep{mueller}.


\begin{figure}[!ht]
\plotfiddle{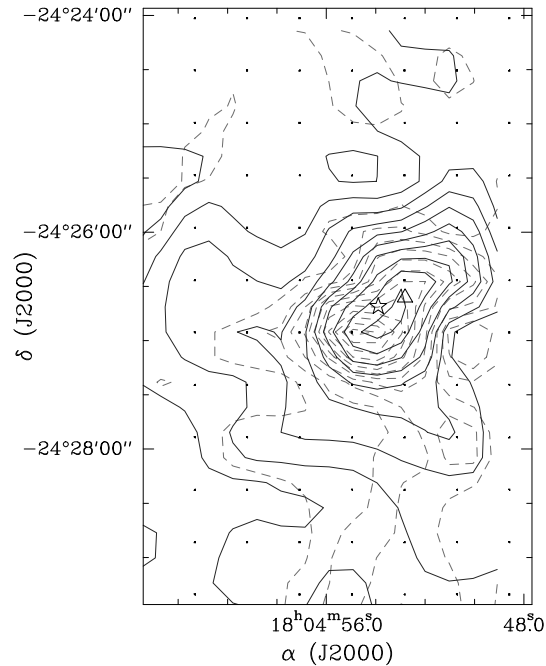}{240pt}{-90}{50}{50}{-130}{270}
\caption{CO 2--1 map of M8\,E from \citet{zhang}. Solid contours denote
blueshifted emission, and dashed contours denote redshifted CO. The star
marks the position of the IRAS source (M8E-IR) and the triangle
denotes the {\sc UCHii} region (M8E-Radio).}
\label{fig-co21}
\end{figure}

As IRAS 18018--2426, M8\,E has been extensively studied as part of a
programme to identify candidate high-mass protostars. It is classified as
`Low' \citep[e.g.][]{beltran}, meaning that its IRAS colours are not
consistent with those of {\sc UCHii} regions, even though it does contain
an {\sc Hii} region with electron density at least 5000~cm$^{-3}$
\citep{molinari98}, which is consistent with the ISO data. Based on NH$_3$
emission, the kinetic temperature of the molecular gas is estimated at
31~K \citep{molinari}, similar to the 29~K estimated from CO lines
\citep{thesispaper}. The clump has also been detected in HCN 1--0
\citep{stclair} and CS 7--6 \citep{plume}. No H$_2$O masers have been
detected towards M8\,E, but OH maser emission has been detected at
1.665~GHz \citep{cohen}, along with methanol masers at 44~GHz and 133~GHz
\citep{slysh,koganslysh}.

The high-velocity molecular gas around M8\,E has been mapped in CO 2--1
\citep[][see Fig.~\ref{fig-co21}]{mmh,zhang}: The red- and blueshifted gas masses are
offset from one another by about 15\arcsec, which suggests a bipolar
outflow with a dynamical timescale of 10$^4$ years \citep{mitch92}.
\citet{mmh} studied the high-velocity molecular gas by the IR absorption of
ro-vibrational transitions, tracing younger
\citep[100 year-old,][]{mitch88}, hotter material. On this evidence,
\citeauthor{mmh} suggest that M8E-IR may be a FU Orionis-type object.
The shortwards end of the ISO data (which should be dominated by M8E-IR)
suggests a B0 star surrounded by a disk \citep{whiteiso}; the best-fitting
models to mid-IR interferometric visibilities \citep{linz} are composed of
a bloated central star (10--15\,M$_\odot$, equivalent to an early B star)
with a small to non-existent disc ($<$50\,AU) surrounded by an envelope
with bipolar cavities. \citeauthor{linz} point out that the massive
accretion events causing FU~Orionis-like outbursts could also cause the
bloating of the central star.

\begin{table}[tbh]
\label{tab-m8e}
\smallskip
\begin{center}
{\small
{Table 6. \hspace{4mm} Stellar sources in M8\,E}\\
\smallskip
\begin{tabular}{lcc}
\tableline
\noalign{\smallskip}
Source  & R.A. (J2000.0) & Dec. (J2000.0) \\
\noalign{\smallskip}
\tableline
\noalign{\smallskip}
Mid-IR source\tablenotemark{a} & 18:04:52.7 & --24:26:41 \\
M8E-Radio\tablenotemark{b,c}   & 18:04:52.8 & --24:26:36 \\
M8E-IR\tablenotemark{b}        & 18:04:53.3 & --24:26:42 \\
S85-4\tablenotemark{b}         & 18:04:53.3 & --24:26:15 \\
S85-3\tablenotemark{b}         & 18:04:53.7 & --24:26:59 \\
S85-2\tablenotemark{b}         & 18:04:53.7 & --24:26:22 \\
S85-1\tablenotemark{b}         & 18:04:54.1 & --24:26:24 \\
\noalign{\smallskip}
\tableline
\noalign{\smallskip}
\multicolumn{3}{l}{$^a$ \citet{linz};~~$^b$ \citet{s85}} \\
\multicolumn{3}{l}{$^c$ Dec.~is misprinted in the original paper} \\
\end{tabular}
}
\end{center}
\end{table}
\stepcounter{table}

\section{Other Candidate Star-Forming Regions}

\label{sec-sfrs}

\subsubsection{M8\,SE3/IRAS\,18014--2428}

Along with M8\,E, this IR source lying within the M8\,SE3 clump, has been
extensively studied as a candidate high-mass protostar. Like M8\,E, it is
classified as a `Low'-type source, is detected in NH$_3$ emission, giving
a kinetic temperature of 27~K (compared with 21~K from CO lines), and is
not associated with an H$_2$O maser \citep{molinari}. It is not detected
in the radio continuum and shows no sign of outflow in molecular lines;
however, the implied axis of the HH\,896--7 jet passes very close to the
IRAS source (see Fig.~\ref{fig-hh}). Whether or not IRAS\,18014--2428 is a
protostar, it seems very likely that the M8\,SE3 clump is a site of
ongoing star formation.

\subsubsection{The Central Ridge}

Submillimetre continuum maps of M\,8 (e.g.~Fig.~\ref{fig-submm}) reveal a
ridge of dense gas running N--S, lying between the Hourglass and M8\,E,
including the submillimetre clumps EC\,1--5. The X-ray data
(Fig.~\ref{fig-acis}) show an overdensity of sources near this ridge,
suggesting that there may be star formation associated with it.

\section{Simeis 188}

\label{sec-sim188}

\begin{figure}
\centering
\includegraphics[draft=False,width=0.92\textwidth]{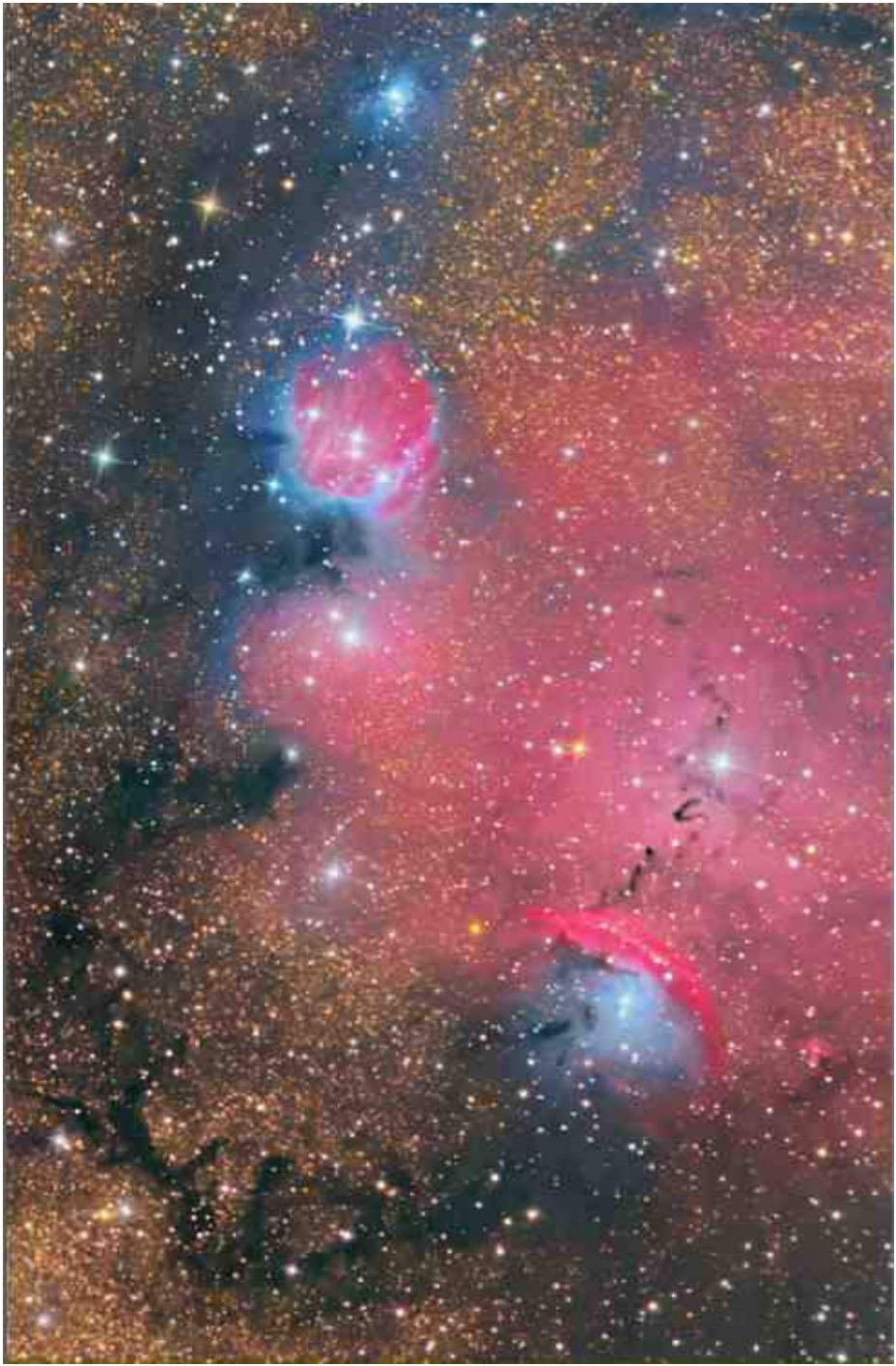}
\caption{The complex of nebulae to the east of M\,8 is known as
Simeis~188 \citep{herbig}, shown here in a three-colour broadband
image. Individual nebulae are identified in
Fig.~\ref{fig-sim188}. North is towards the top left, and east towards
the bottom left; FOV is $\sim 35^\prime\times\sim 55^\prime$; see also
Fig.~\ref{fig-sim188}. Courtesy Tony Hallas.}
\label{fig-sim188-colour}
\end{figure}

\begin{figure}[thb]
\plotone{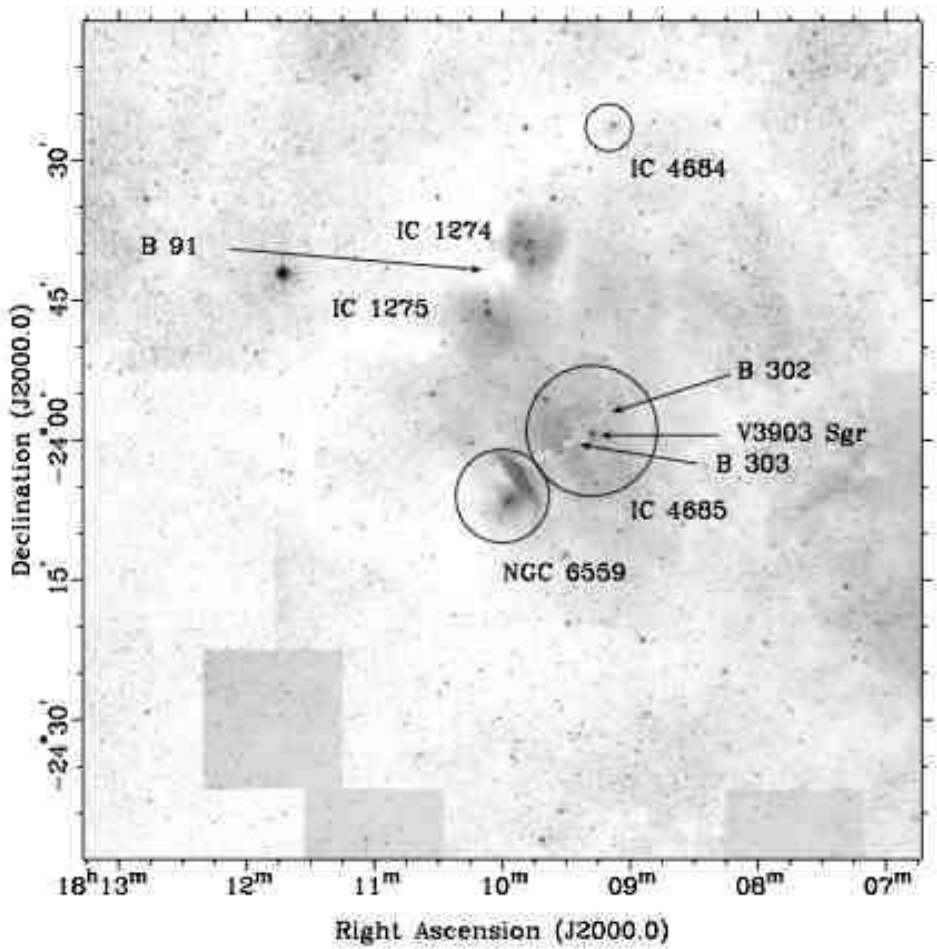}
\caption{DSS2 red image of the Simeis~188 region. The loose cluster
Collinder\,367 is found within the diffuse {\sc Hii} region IC\,4685,
centred on the O-star V\,3903~Sgr.}
\label{fig-sim188}
\end{figure}

About a degree to the east of the Lagoon Nebula,
{Simeis~188}\footnote{The designation comes from
the catalogue of \citet{simeis} developed at the Simeiz Observatory in
Ukraine; these objects are traditionally spelled Simeis.} comprises
emission and reflection nebulae and dark clouds
\citep[Fig.~\ref{fig-sim188-colour}; also][]{barnard92}, and lies near a
loose open cluster, Collinder\,367, located within the {\sc Hii} region
IC\,4685. The nebular features are identified in the annotated red-light
DSS image (Figure~\ref{fig-sim188}) and their positions listed in
Table~7 IC\,4685 is a large diffuse nebula centred on
V\,3903~Sgr (HD\,165921), an eclipsing binary comprising two main-sequence
O~stars (O7\,V\,+\,O9\,V), also discussed below \citep{vaz}. On the
southeastern edge of IC\,4685, NGC\,6559 is a bright-rimmed cloud, with
the rim running NE--SW
(see Figs.~\ref{fig-sim188-colour} \& \ref{fig-sim188}), and lying close to
IRAS\,18068--2405 \citep{sugitani,02ogura}. The northeastern edge of
Simeis~188 is dominated by two bright emission nebulae: IC\,1274 in the
north, illuminated by two stars, and IC\,1275; these two are separated by
the dark cloud B\,91. North of IC\,1274 lies IC\,4684, a very small
reflection nebula around the 9th magnitude B3 star HD\,165872. There are
two more Barnard dark nebulae lying in front of IC\,4685: B\,302 and
B\,303, just NW and SE of V\,3903~Sgr, respectively. They seem to be
components of a long thin dust lane lying across and in front of the
emission nebula.

\begin{table}[tbh]
\label{tab-sim188}
\smallskip
\begin{center}
{\small
{Table 7. \hspace{4mm} Nebulae in Simeis 188}\\
\smallskip
\begin{tabular}{lcc}
\tableline
\noalign{\smallskip}
Nebula & RA (J2000.0) & Dec (J2000.0) \\
\noalign{\smallskip}
\tableline
\noalign{\smallskip}
IC 4684  & 18:09:06 & --23:25 \\
IC 4685  & 18:09:18 & --23:59 \\
IC 1274  & 18:09:30 & --23:44 \\
IC 1275  & 18:10:00 & --23:50 \\
NGC 6559 & 18:10:00 & --24:06 \\
B 302    & 18:09:14 & --23:58 \\
B 303    & 18:09:29 & --24:00 \\
B 91     & 18:10:08 & --23:42 \\
\noalign{\smallskip}
\tableline
\noalign{\smallskip}
\multicolumn{3}{l}{Nebula data from \citet{uranometria}} \\
\end{tabular}
}
\end{center}
\end{table}
\stepcounter{table}

Collinder\,367 is older than NGC\,6530, but still fairly young, and lies
at a similar distance \citep[$\sim$7~Myr and $\sim$1.2~kpc;][]{kharchenko}.
V\,3903~Sgr is much younger, with an age (2~Myr) and distance similar to
NGC\,6530 \citep{vaz}, and may represent a later generation of star
formation. There is some evidence that this complex could be related to
the Lagoon: \citet{herbst} found that the various stars illuminating the
nebulosity show reddening $R=4.2$, quite similar to the abnormal reddening
found towards M\,8, and the nebulosity around M\,8 extends all the way out
to Simeis\,188 \citep[Fig.~\ref{fig-wide}; also][]{barnard08}.

\begin{figure}
\plotfiddle{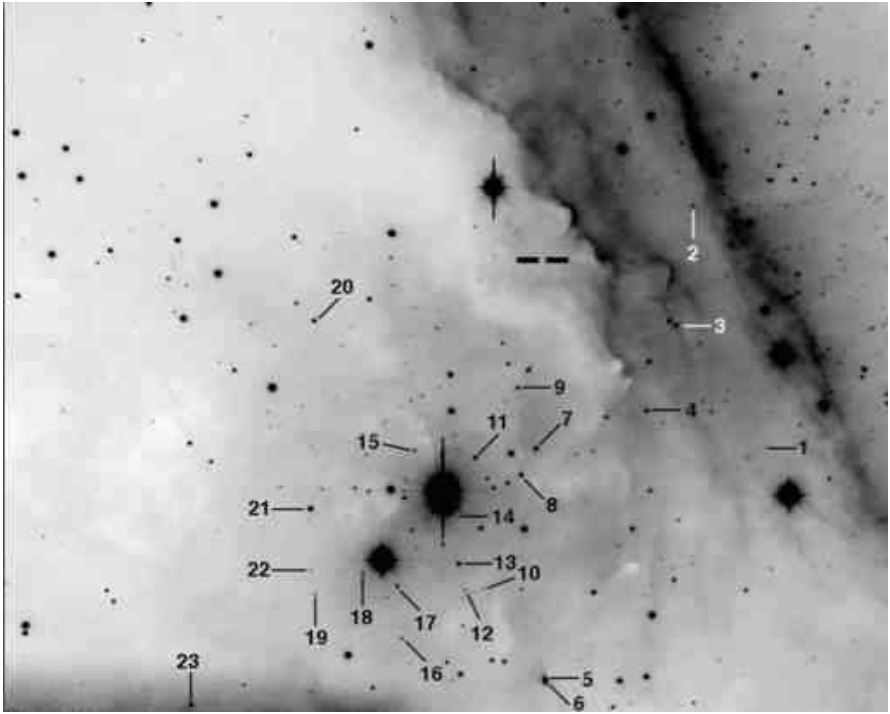}{4.in}{0}{60}{60}{-190}{10}
\caption{H$\alpha$ stars in the bright-rimmed cloud NGC\,6559 (BRC\,89).
North is up and East is to the left; FOV is $\sim 4^\prime\times\sim 6^\prime$;
the thick tick marks denote the position of IRAS\,18068--2405. From
\citet{02ogura}.}
\label{fig-halpha}
\end{figure}

\citet{herbig} found 6 H$\alpha$ emission stars (LkH$\alpha$\,125--130) in
the Simeis~188 region, of which he suggests that LkH$\alpha$~125 is unlikely
to be associated with the nebulae. Based on the finding charts and
instructions from that paper, positions of all 6 stars have been taken from
the Digitised Sky Survey (or 2MASS in the case of LkH$\alpha$~130), and are
given in Table~8. \citet{02ogura} found a further 23
H$\alpha$ emission stars (but no Herbig-Haro objects): These fainter
stars (Fig.~\ref{fig-halpha}) are probably T~Tauri stars,
whereas the 6 stars from \citet{herbig} are more likely to be HAeBes.

\begin{table}[tbp]
\label{tab-pms_sim188}
\smallskip
\begin{center}
{\small
{Table 8. \hspace{4mm} H$\alpha$ emission stars in Simeis 188, from \citet{herbig}}\\
\smallskip
\begin{tabular}{lcc}
\tableline
\noalign{\smallskip}
Star & R.A.\,(J2000.0) & Dec.\,(J2000.0)  \\
\noalign{\smallskip}
\tableline
\noalign{\smallskip}
LkH$\alpha$\,125 & 18:07:58.3 & --23:33:38\tablenotemark{a} \\
LkH$\alpha$\,126 & 18:09:23.6 & --23:27:46\tablenotemark{a} \\
LkH$\alpha$\,127 & 18:09:35.7 & --23:25:19\tablenotemark{a} \\
LkH$\alpha$\,128 & 18:09:45.5 & --23:38:03\tablenotemark{a} \\
LkH$\alpha$\,129 & 18:09:46.1 & --23:38:52\tablenotemark{a} \\
LkH$\alpha$\,130 & 18:09:47.3 & --23:38:43\tablenotemark{b} \\
\noalign{\smallskip}
\tableline
\noalign{\smallskip}
\multicolumn{3}{l}{$^a$ from WCS of DSS image} \\
\multicolumn{3}{l}{$^b$ from WCS of 2MASS $J$-band image} \\
\end{tabular}
}
\end{center}
\end{table}
\stepcounter{table}

\section{The Structure and Evolution of the Lagoon Nebula}

\label{sec-discussion}

\begin{figure}
\plotone{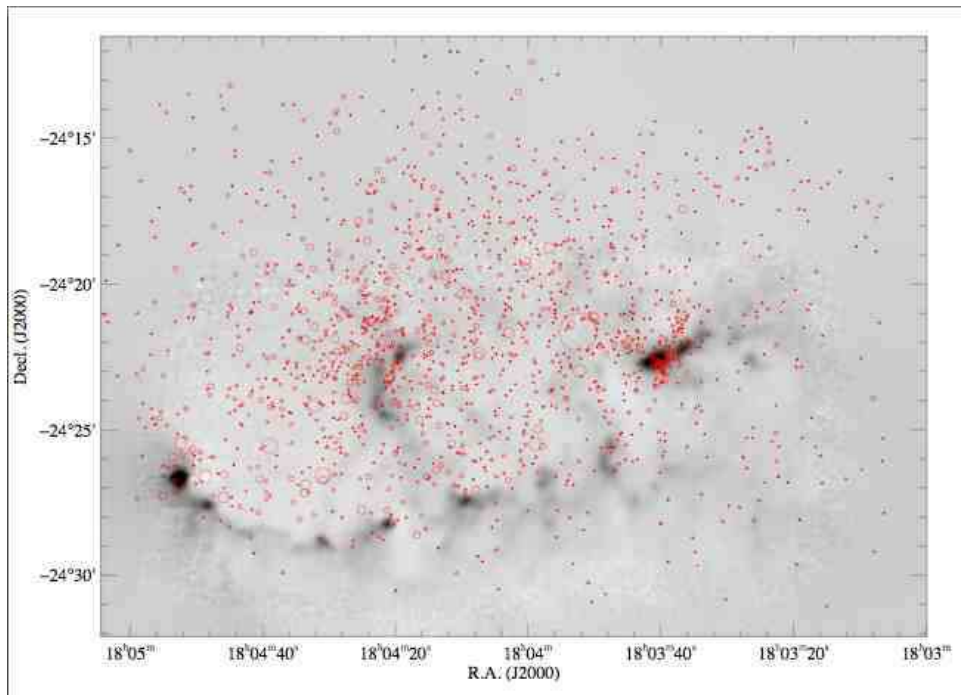}
\caption{{\it Chandra} X-ray sources overlaid on the sub-mm continuum
structure of the Lagoon Nebula, from \citet{gagne}.}
\label{fig-acis}
\end{figure}

\begin{figure}
\plotone{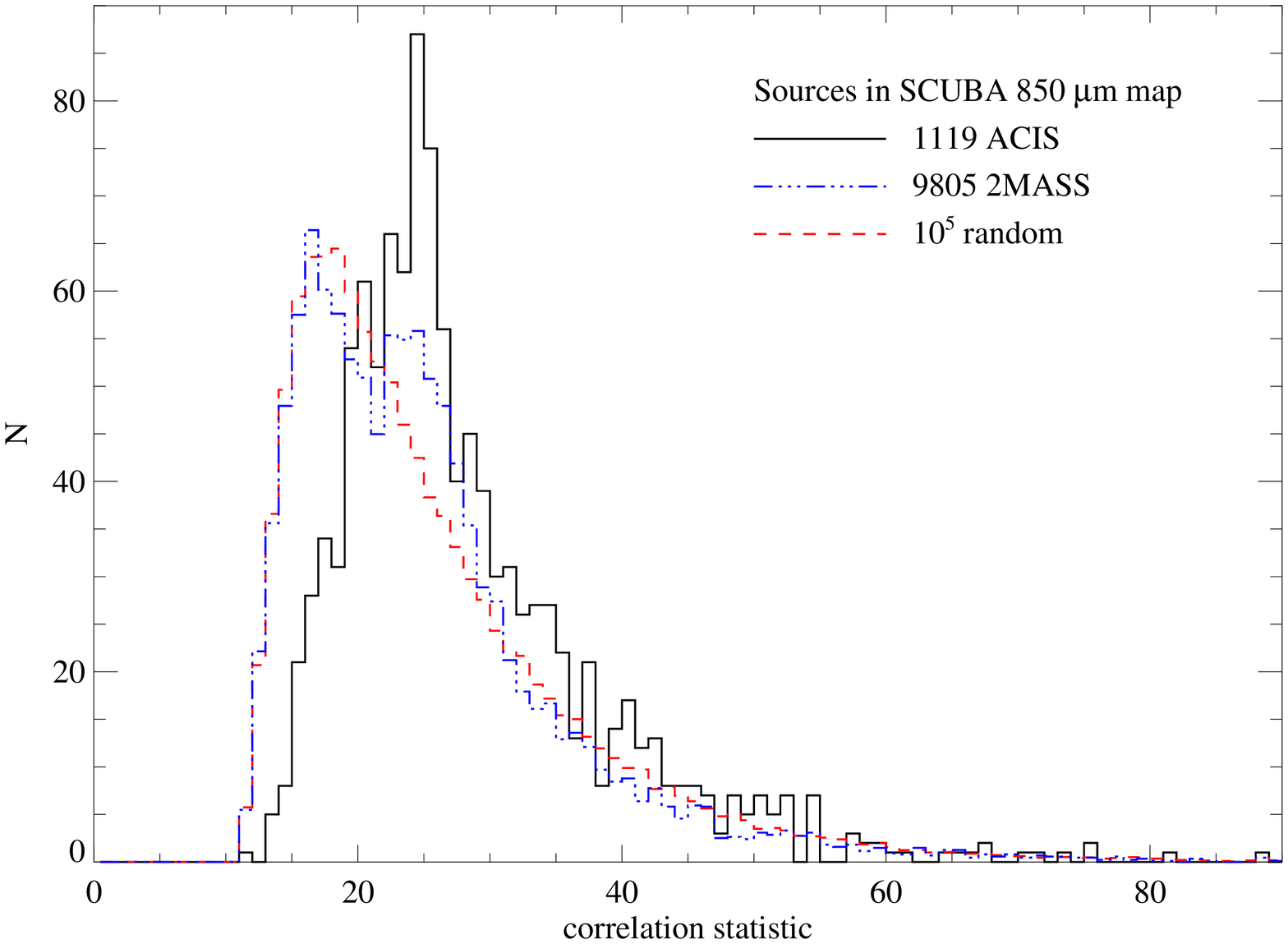}
\caption{Spatial correlation between 850~\micron\ emission and
{\it Chandra} X-ray sources, 2MASS sources, and a random distribution,
from \citet{gagne}.}
\label{fig-acis-correl}
\end{figure}

\subsubsection{The Structure of Star Formation}

Combining the {\it Chandra} images towards the Hourglass \citep{gagne}
and NGC\,6530 \citep{dam} yields a catalogue of 1482 X-ray sources
spanning most of the region observed at 450 and 850~\micron\ by
\citet{thesispaper}. Figure~\ref{fig-acis} suggests that X-ray sources
tend to cluster near sub-mm emission cores, which is confirmed by
Figure~\ref{fig-acis-correl}: This shows the spatial correlation of the
850~\micron\ flux to the location of each of the 1119 X-ray sources lying
within the SCUBA map, and the same statistic for the 9805 2MASS sources in
the map and for a set of $10^5$ random sources. The distributions show that
the X-ray source locations are non-random and correlated with 850~\micron\
flux, as are about 20\% of the 2MASS sources. In particular,
Fig.~\ref{fig-acis} shows strong clustering around (but slightly offset from)
M8\,E (near 18:04:54, --24:26:30), the central ridge (near 18:04:20, --24:23),
and Herschel~36 and the Hourglass (near 18:03:40, --24:23). In the central
ridge (M8\,EC1--5), three lines of X-ray sources are seen separated by
$\sim 1\arcmin$ from northeast to southwest. A number of smaller X-ray
clusters appear close to 850~\micron\ cores along the southern rim,
specifically: M8\,SC8, SC1, SE1, SE3, and SE7. \citet{barba} also found
H-H objects associated with M8\,SC8, SE3 and C3.

There are other bright rims and globules in and around the Lagoon, most
of which seem to be due to the action of 9\,Sgr, such as the bright rims
of \citet{sugitani}, the globule near the Hourglass \citep{arias05} and
the elephant trunk found near M8\,E \citep{brand}. These structures, along
with the widespread T~Tauri star population \citep{pris,arias07}, suggest
pervasive star formation in the Lagoon, in addition to the young clusters
found around the Hourglass and M8\,E.

\subsubsection{Star Formation History}

While most authors ascribe an age of a few Myr to NGC 6530, the oldest
element of the Lagoon Nebula, \citet{vda} argue for a much older cluster
(by an order of magnitude), based on the probable cluster members found
in the giant branch of the H-R diagram. These members seem to be worth
investigation: If star formation has really been going on for a few tens of
Myr, and is still ongoing at a significant rate, M\,8 would be a very
long-lived star-forming region. \citeauthor{vda} also suggest, because of
the lack of massive stars on the ZAMS, that massive star formation has
essentially ceased. This may be true of NGC\,6530, but massive star formation
in M\,8 as a whole is not over.

\citet{lightfoot} identified a possible sequence of triggered star
formation in M\,8: NGC\,6530 is the oldest feature, and some of its members
are still ionising the main {\sc Hii} region, NGC~6523, while younger
features are found around the edge of the region, most obviously in the
Hourglass and M8\,E. The O stars in the cluster are also found at the
periphery, while the core contains only B stars and later, and it is not
clear why this should be so: Do the peripheral O stars represent a later
generation of star formation? If so, it is odd that there are no signs of
any O stars corresponding to the $\sim$60 B stars in the core, either as
main-sequence O stars or as post-MS objects. However, the basic picture
of star formation proceeding outwards from the core of NGC\,6530 is
well-supported by evidence of age gradients \citep[e.g.~][]{dam,arias07}
as well as the prevalence of massive young stars, X-ray sources and
H-H objects around the dense molecular cores around the edge of the
Lagoon. There are also signs of star formation elsewhere in M\,8, e.g.~the
sample of candidate Class~I sources concentrated to the northeast of the
Hourglass \citep{dam06}.

\subsubsection{The Structure of the Lagoon Nebula}

\begin{figure}
\plotone{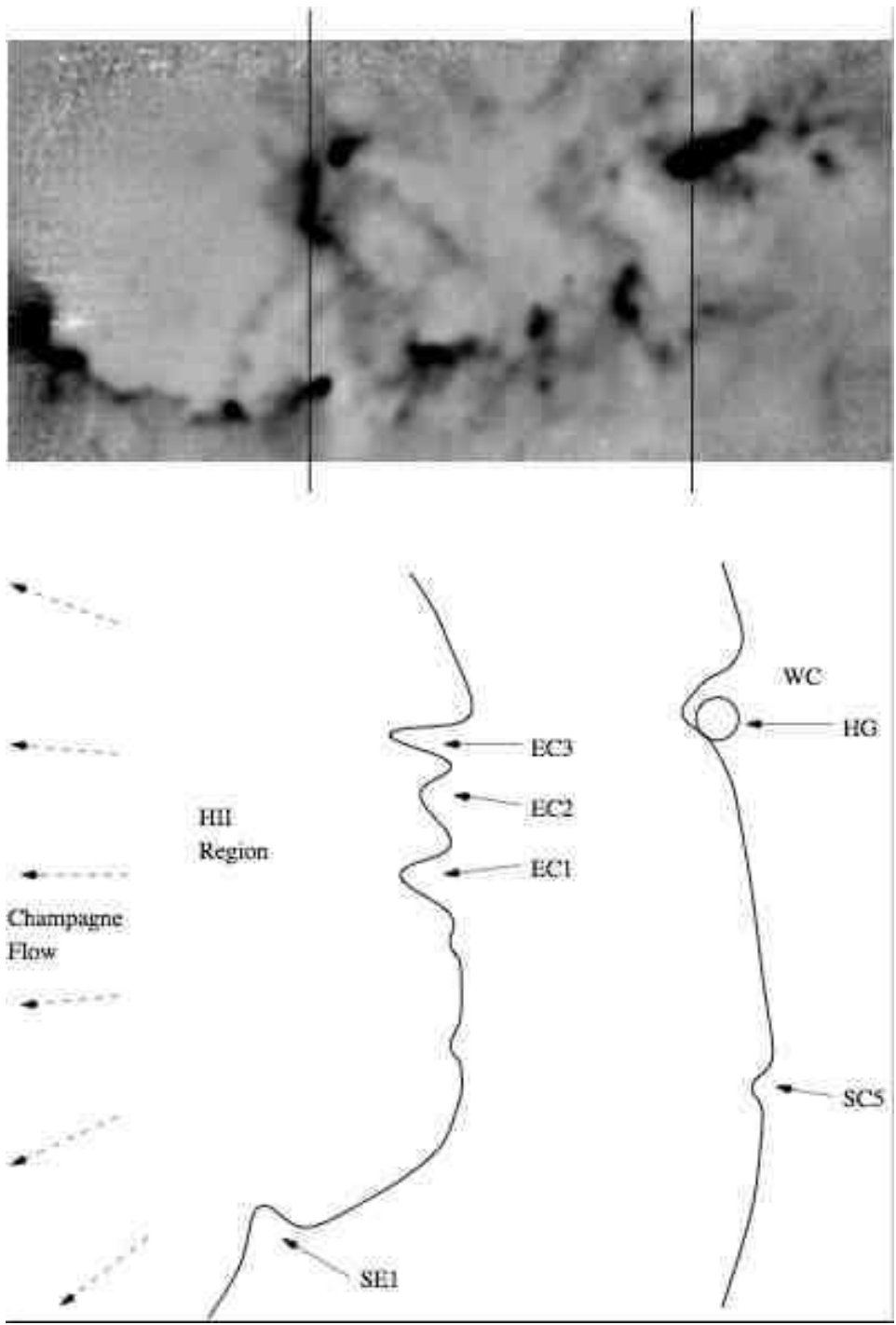}
\caption{Schematic diagram showing a possible structure of the Lagoon
Nebula, in the form of two cuts through the Central Ridge and Hourglass
regions (shown on a $\sim 20^\prime\times\sim 10^\prime$ section of the
850~\micron\ emission map). From \citet{thesis}.}
\label{fig-xsection}
\end{figure}

Compared to the clumps along the southern and southeastern rims
of the Lagoon Nebula, the EC clumps seem to fall off fairly shallowly
on all sides. This suggests that the central ridge may lie behind the
{\sc Hii} region, and we see the structure face-on, rather than the
edge-on view of the southern clumps (see Fig.~\ref{fig-xsection}).
\citet{woodward} argue convincingly that the Hourglass is embedded in
the molecular cloud behind the Lagoon, whereas M8\,E shows a steep fall-off
into the {\sc Hii} region. So we see ongoing star formation both behind
and to the southeast of the ionised gas, suggesting that the ionisation
front is moving away from us and to the south and east, compressing and
warming the molecular gas (giving rise to submillimetre and CO emission)
and presumably triggering star formation, with X-ray emitting YSOs
appearing tightly clustered in the wake of the front.

There are indications of a thin screen of material between us and M\,8,
somewhat blueshifted. This might be the last remnants of the front of
the molecular cloud, excavated by a blister {\sc Hii} region on the
front side of the cloud, and accelerated towards us by the ionised gas.

\section{Closing Remarks}

\citet{bok} commented, at the end of their paper on globules in the Milky
Way, many of them in M\,8, that every one of the globules they had just
described merited further careful study ``with the largest available
reflecting telescopes''. We believe that this advice is still valid, 60
years later. New observing facilities and techniques at many different
wavelengths are giving us new opportunities to understand this region:
The impact of the latest X-ray observations is immense, since it makes
it much easier to disentangle young stars from the background. The wider
availability of large telescopes, allowing photometry and spectroscopy of
faint stars, enables proper classification of the lower-mass PMS population
\citep[e.g.][]{arias07}. {\it Spitzer} data (e.g.~Fig.~\ref{fig-irac})
offer further opportunities to select sources of interest out of the
crowded background. Using time-domain photometry, asteroseismology allows
detailed modelling of PMS stars in NGC\,6530 \citep{daveg}.
Aperture-synthesis observations at millimetre- and submillimetre-wavelengths
may be used to compensate for the effects of distance, in order to search
for protostars in the molecular cores around the Nebula; mid-IR
interferometry is already yielding new insights into the massive young
stars in M8\,E \citep{linz}.

\vspace{0.5cm}

{\bf Acknowledgements.}  We are most grateful to the referees, Julia
Arias and Rodolfo Barb\'a, whose careful reading and comments have
improved this work. We also thank Steve Rodney for his work on the
bibliography, Qizhou Zhang, Eric Mamajek, Phil Castro and Nathan Mayne
for helpful discussions, and Bo Reipurth for his editorial work.  We
thank Gerald Rhemann, Richard Crisp, Jean-Charles Cuillandre, and Tony
Hallas for use of Figures 1, 4, 7b, and 13, respectively.
NFHT gratefully acknowledges financial
support from the University of Exeter DVC (Resources) Discretionary Fund
and from the European Commission (grant MIRG-CT-2006-044961), and thanks
the crew of Amundsen-Scott South Pole Station, where an early draft was
written. This review has made use of: NASA's Astrophysics Data System;
the SIMBAD database, operated at CDS, Strasbourg; the {\it Skyview}
facility located at NASA's Goddard Space Flight Center; the WEBDA
database, operated at the Institute for Astronomy at the
University of Vienna.

\end{document}